%% file: main.tex
\documentclass[acmlarge,screen,authorversion,nonacm]{acmart}

\usepackage{booktabs}
\usepackage{listings}
\usepackage{tabularx}
\usepackage{amsmath,amsfonts}
\usepackage{algorithmic}
\usepackage{graphicx}
\usepackage{textcomp}
\usepackage{xcolor}
\usepackage{enumitem}
\usepackage{url}
\usepackage{hyperref}
\usepackage{xspace}

\newcommand{\todo}[1]{}
\renewcommand{\todo}[1]{{\color{red} {#1}}} 

\newcommand{\defectsFourJ}{Defects4J\xspace}
\newcommand{\deepseekROneDist}{DeepSeek R1 (dist.)\xspace}

\begin{document}

\title{Exploring Generalizable Automated Program Repair with Large Language Models}


\author{Viola Campos}
\email{viola.campos@hs-rm.de}
\orcid{0009-0004-0188-6154}
\affiliation{%
  \institution{RheinMain University of Applied Sciences}
  \city{Wiesbaden}
  \country{Germany}
}

\author{Ridwan Shariffdeen}
\email{shariffdeenr@acm.org}
\orcid{0000-0001-5409-4864}
\affiliation{%
  \institution{SonarSource}
  \city{Singapore}
  \country{Singapore}
}

\author{Adrian Ulges}
\email{adrian.ulges@hs-rm.de}
\orcid{0009-0001-1915-2464}
\affiliation{%
  \institution{RheinMain University of Applied Sciences}
  \city{Wiesbaden}
  \country{Germany}
}

\author{Yannic Noller}
\email{yannic.noller@acm.org}
\orcid{0000-0002-9318-8027}
\affiliation{%
  \institution{Ruhr University Bochum}
  \city{Bochum}
  \country{Germany}
}

\renewcommand{\shortauthors}{Campos et al.}

\begin{abstract}

Automated Program Repair (APR) proposes bug fixes to aid developers in maintaining software. The state of the art in this domain focuses on LLMs, leveraging their strong capabilities to comprehend specifications in natural language and to generate program code. However, despite the APR community's research achievements and industry deployments, APR still cannot generalize broadly.
In this work, we present an intensive empirical evaluation of LLMs' capabilities in APR. We evaluate a diverse set of 13 recent open and closed models. In particular, we explore language-agnostic repair by utilizing benchmarks for Java, JavaScript, Python, and PHP.
Besides the generalization across languages and levels of patch complexity, we also investigate the effects of fault localization (FL).
Our key results include: (1) Different LLMs tend to perform best for different languages, which makes it hard to develop cross-platform, single-LLM repair techniques. (2) Combining models by pooling repairs adds value with respect to uniquely fixed bugs, so a committee of expert models should be considered. (3) Under realistic assumptions of imperfect FL, we observe significant drops in accuracy from the usual practice of using perfect FL. Our insights will help develop reliable and generalizable APR techniques and evaluate them in realistic and fair environments.
\end{abstract}




\maketitle

\input{intro}
\input{related}
\input{methodology}
\input{results}
\input{discussion}

\input{conclusion}

\section{Data Availability}
The data that support the findings of this study, including all our results and the generated patches, are openly available in our supplemental material:
\url{https://figshare.com/s/947fd7030f10a67a1c9f}.
Upon acceptance, we will update the artifact with a reproduction package including the scripts for prompting the LLMs as well for the data analysis, and make the material accessible via a DOI.

\input{disclaimer}


\bibliographystyle{ACM-Reference-Format}
\bibliography{references}


\end{document}

%% file: intro.tex
\section{Introduction}

Automated Program Repair (APR)~\cite{APRieeeSoftware2021,gao2022programrepair} has gained significant attention in the last few years. It attempts to repair software bugs automatically, and hence, promises great support for software practitioners in maintaining code.
%
APR research in the last decade has focused on search-based, semantics-based, template-based, and learning-based methods. Search-based techniques~\cite{genprog} apply evolutionary code mutations to change the program to pass a test suite. Semantics-based techniques~\cite{semfix} use a test suite to extract specifications as constraints via, e.g., symbolic execution. These constraints then drive a program synthesis to generate a repair, e.g., expressions or whole statements. Template-based techniques~\cite{tbar} use pre-defined patch templates that are either manually crafted for a specific domain or mined from large code repositories. Finally, learning-based techniques~\cite{iter} attempt to directly learn how to generate patches based on large sets of historic fixes. 

Despite these research achievements in APR and existing industrial deployments~\cite{bloomberg2024,inferfix,facebook2023}, there are still issues with generalizing and benchmark overfitting.
The first international competition on APR~\cite{aprcomp2024} in 2024 using the Cerberus framework~\cite{DBLP:conf/icse/ShariffdeenMNR23} has shown that current techniques are not yet able to reliably provide software patches, i.e., they failed to work in realistic settings and generalize beyond the community benchmarks. For example, in the \textit{Functional Error} track for Java, the best tool was able to correctly fix only $5$ out of $50$ previously unseen bugs (i.e., 10\% fix rate) and many techniques were unable to provide any correct patches. The benchmarks had been curated with new and unforeseen repair tasks, and the tools had been evaluated in a uniform environment enforcing strict resource constraints.

Machine Learning (ML) techniques, specifically Large Language Models (LLM), play a significant role in the SE and APR community's most recent research efforts. 
Due to their pre-training on large amounts of open-source code and due to their enormous capacity, LLMs have demonstrated problem-solving skills over a wide range of tasks~\cite{zhao_survey_2023}.
In APR, recent works~\cite{alpharepair,chatrepair} have shown that LLMs can be used to generate repairs and build SE agents \cite{autocoderover,specrover,yang2024swe}, which attempt to mimic the work of software engineers. Nevertheless, most of the current LLM-based APR approaches use LLMs in a rather ad-hoc strategy (e.g., ~\cite{alpharepair,fitrepair,tare2023,hossain2024}), and while some existing works study the usage of LLMs in a more systematic way (e.g.,~\cite{huang2023empirical,silva2024repairllama,Jiang2023}), they all fall short on at least one of the following points: (i) the investigation is based on one specific LLM only, (ii) no/limited exploration of various prompting strategies/designs is conducted, in particular regarding repair ingredients (like information about failing test cases, bug reports/descriptions, error messages, etc.), (iii) a perfect Fault Localization (FL) is assumed to be given, which represents an unrealistic assumption in a practical setting, and (iv) only one specific benchmark with one programming language is covered, e.g., \defectsFourJ~\cite{DBLP:conf/issta/JustJE14}.


To address the above issues and to provide a comprehensive overview of the current research landscape of APR, this work presents a systematic empirical evaluation with state-of-the-art \textit{closed} and \textit{open} LLMs on multiple benchmarks \textit{across} programming languages.
Hereby, we refer to \textit{open} models as those with publicly available weights\footnote{Note that, according to our definition, open models may not necessarily come with open-source licenses, and that -- unlike their weights -- their training code and/or data may remain undisclosed.}, whereas \textit{closed} LLMs are proprietary models whose weights, training data, and model architectures have not been publicly released. 
Particularly, we compare closed models like Claude 3.7 Sonnet, Gemini 1.5 Pro, OpenAI o3-mini, and OpenAI GPT-4o
with open ones like Llama 3.3, CodeLlama, and DeepSeek Coder on benchmarks for Java (\defectsFourJ~\cite{DBLP:conf/issta/JustJE14}), JavaScript (BugsJS~\cite{bugsjs2023}), Python (BugsInPy~\cite{DBLP:conf/sigsoft/WidyasariSLQPTT20}), and PHP (BugsPHP~\cite{DBLP:conf/msr/PramodSTSW24}).
In total, we generated and evaluated $\approx 195,000$ patches in four programming languages and fixed $542$ out of $712$ bugs.

We study the following aspects: (1) The question of whether certain LLMs display language-agnostic repair capabilities.
(2) The influence of different information sources in the prompt (specifically, localization and test errors). 
(3) The robustness of LLMs to the quality of Fault Localization (FL). To do so, we systematically evaluate LLMs' APR accuracy for line-level vs function-level localization, and when combined with a real-world spectrum-based automated FL.
(4) The performance across different patch complexity levels (i.e., single-line, single-hunk, multi-hunk).

Based on our experiments, current LLMs do not generalize well across programming languages for the task of program repair. In particular, we observed that different LLMs perform well on different benchmarks. 
Similar to others~\cite{DBLP:journals/corr/abs-2404-05520}, we observed that including information about the failing test case significantly boosted the repair performance for all models.  
We found LLMs to show good robustness when repairing bugs of higher complexity.
However, notably, our comparison between perfect and automated FL techniques showed a drastic decline in repair capabilities if the buggy function cannot reliably be identified.
Overall, our findings provide fundamental and comprehensive understanding of the current capabilities of LLMs with respect to the various aspects of APR, which is beneficial for our community (e.g., for designing future research efforts) and for practitioners (e.g., for selecting appropriate LLMs for their repair use case).
In summary, we make the following core contributions:
\begin{itemize}
    \item We provide an extensive evaluation of state-of-the-art (open and closed) LLMs' program repair capabilities, including multiple models, programming languages, prompt variations, patch complexities, and fault localization methods.
    \item To the APR practitioner, our main take-away results are that (1) different LLMs are suited best for different languages, suggesting language-specific model committees, (2) test information appears to be a key signal for LLM-based APR, and (3) open models appear to be catching up to their closed counterparts.
    \item To the future APR researcher, our study shows that (1) fault localization is a potential bottleneck for APR pipelines, which suggests a strong focus on this aspect. 
    (2) APR accuracy generalizes poorly, with different models reacting differently when varying programming language or prompt information; and 
     (3) combining different LLMs improves results across languages and offers interesting directions for future work.
    We conclude that APR pipelines need to be benchmarked carefully and broadly.    
  To support such a holistic evaluation of future progress in APR, we open-source our experimental setup and results for other researchers. 
\end{itemize}


%% file: related.tex
\section{Background and Related Work}



\subsection{LLM based Program Repair}
Numerous APR techniques have been studied in the last decade, ranging from the traditional search-based repair~\cite{genprog}, constraint-based repair~\cite{semfix}, template-based repair~\cite{tbar}, and learning-based repair~\cite{prophet}. More recently, with the advancement of LLMs for code generation, various techniques~\cite{alpharepair, chatrepair, iter} have been explored by the research community to utilize LLMs for the program repair problem. LLM-based repair techniques transform the program repair problem into a code generation problem. These techniques can be broadly classified as model training~\cite{silva2024repairllama, fitrepair}, zero-shot learning~\cite{alpharepair} and few-shot learning~\cite{Tang2024, inferfix, thinkrepair}. Silva et al.~\cite{silva2024repairllama} explored how different input and output representations yield optimized results using LLMs for program repair tasks. Xia et al.~\cite{chatrepair} studied the effect of test diagnostic feedback for LLM-driven program repair. Lin et al.~\cite{Lin2024} studied the impact of repair granularities encountered in program repair tasks. More specifically, they examined how expression-level and token-level repairs differ for different LLMs. 

In our work, we incorporate these different findings and techniques into an extensive study with 13 state-of-the-art LLMs and evaluate the generalizability across different languages. We also evaluate the repairability of the latest models by varying the information we provide (i.e., test diagnostic) and varying the granularity of the task (function level vs statement level).


\subsection{Empirical Studies of Program Repair}
Several empirical studies have been conducted to evaluate program repair techniques in their efficacy~\cite{Liu2020}, performance~\cite{Zhong2023}, trust~\cite{Noller2022}, and overfitting~\cite{Durieux2019,Ye2019} aspects. Ye et al.~\cite{Ye2019} conducted an empirical study using 10 pre-LLM program repair tools for Java programs and found overfitting to the respective test suites to be an issue. Similarly, Durieux et al.~\cite{Durieux2019} conducted a large-scale evaluation using 11 APR tools to fix 2,141 Java programs in five different repair benchmarks. Their experiments showed that repair tools do not generalize across benchmarks, highlighting the problem of benchmark overfitting in repair techniques. Liu et al.~\cite{Liu2020} conducted another study on the efficiency of repair tools using 16 Java repair tools. Their study highlighted the performance inefficiencies in repair tools, which predominantly generated a large number of incorrect patches despite their overall repair capabilities. Noller et al.~\cite{Noller2022} performed a small-scale experiment using five program repair tools and 60 C programs. Their findings reported that different experiment configurations lead to different results, highlighting the importance of fair comparison under a uniform setup. A similar study was conducted by Zhong et al.~\cite{Zhong2023} for learning-based repair tools, highlighting the importance of a unified experimental setup with respect to the training data and evaluation metrics used for comparisons.

More recently, Xia et al.~\cite{Xia2023} conducted an extensive evaluation using 9 LLMs and five repair benchmarks covering multiple languages under different repair configurations. Their findings highlighted the benefits of using LLMs for APR, with respect to the quality of the generated patches and the performance improvements compared to non-LLM-based APR tools. Several other studies~\cite{Jiang2023, huang2023empirical} investigated fine-tuned coding models for program repair using multiple benchmarks, showing they can outperform non-LLM-based repair tools. Ouyang et al.~\cite{Ouyang2024} studied 11 repair tools including traditional techniques, NMT-based, and LLM-based repair tools. Their findings further emphasized the benefits of using LLM-based repair tools, which showed less susceptibility to overfitting and better generalization across benchmarks. Unlike existing work, this study performs an extensive analysis using the most recent 13 LLMs including the most advanced reasoning models across various experiment configurations. 
We focus on the generalizability of LLMs across different languages, repair configurations and patch complexities. Additionally, we revisit the impact of accurate fault localization on program repair with respect to the usage of the most advanced reasoning LLMs.




    



%% file: methodology.tex
\section{Study Methodology} 

\lstdefinelanguage{Prompt}{
  morekeywords={def, return, if, else, for, while, class, import},
  sensitive=true,
  morecomment=[l]{\/\/},
  morestring=[b]",
}

\lstset{
  basicstyle=\ttfamily\scriptsize,
  breaklines=true,
  frame=single,
  keywordstyle=\color{blue},
  commentstyle=\color{gray}\itshape,
  stringstyle=\color{teal},
  showstringspaces=false,
 escapeinside={(*}{*)}, 
  language=Prompt
}

The primary objective of this work is to explore the feasibility of APR using current LLMs. We first discuss the four research questions addressed in our study (Section \ref{sec:rqs}), followed by the $13$ LLMs we evaluate (Section \ref{sec:models}) and the $4$ benchmark datasets in different programming languages (Section \ref{sec:benchmarks}). 
We outline further technical details -- the evaluation protocol, technical setup and prompt design -- in Sections \ref{sec:metrics} to \ref{sec:prompt-design}.

\subsection{Research Questions} \label{sec:rqs}
To assess the accuracy of LLM-based program repair across different influencing factors, we formulate the following research questions:
\begin{enumerate}[label=\textit{RQ\arabic*},start=1,leftmargin=1.2cm]
\item How well does LLM-based APR generalize across different programming languages?
\item How is LLM-based repair impacted by fault localization (FL), particularly when paired with a non-ML localization of faulty code segments?
\item How does LLM-based repair perform across varying levels of patch complexity?
\item How do open LLMs compare to closed ones in the context of APR?
\end{enumerate}

\paragraph*{RQ1 -- Generalization across Programming Languages}
To answer this question, we apply LLM-based APR on repair benchmarks for Java, Python, PHP, and JavaScript and assess the quality of the generated patches.
As part of this question, we also investigate whether one model appears superior across all programming languages, or whether different models work best on specific languages (which would suggest a language-specific model selection).
To investigate this question, we use two prompt designs (see Section~\ref{sec:prompt-design}); (1) a basic prompt including only the code passage to be repaired, and (2) an extended prompt which also includes additional information from failing tests' error messages.

\paragraph*{RQ2 -- Impact of Fault Localization (FL)}
Traditionally, APR pipelines feature a fault localization (FL) step (in which a faulty code subsegment is identified) and a repair step (in which the subsegment is fixed). LLMs could either follow this approach and take localization hints from external systems such as FLACOCO~\cite{Silva2021FLACOCOFL} using spectrum-based fault localization~\cite{DBLP:journals/corr/SouzaCK16}, or operate in an end-to-end fashion, fixing a given chunk of code without explicitly localizing faulty subsegments. So far, there are no conclusive research results as to what extent LLMs benefit from localization, with a lot of research assuming perfect localization to be given~\cite{alpharepair, chatrepair, hossain2024,huang2023empirical,silva2024repairllama,Jiang2023}, others focusing on end-to-end operation~\cite{DBLP:journals/corr/abs-2409-18952}, and only few recent comparisons of both with inconclusive results so far~\cite{silva2024repairllama,xu2024aligning}. 
To this end, we investigate whether current LLMs benefit from localization signals at code line granularity.

\paragraph*{RQ3 -- Patch Complexities}
To better understand LLMs' repair capabilities, we explore our results with respect to various patch complexities: single-line, single-hunk, and multi-hunk bugs (i.e., those which span multiple, separate locations within the same function). While we can expect that LLMs can repair single-line bugs, it is still unclear how they perform on multi-hunk ones. 

\paragraph*{RQ4 -- Open vs Closed LLMs}
In our last question, we focus on the differences in APR accuracy between open LLMs and closed LLMs.
%
While closed models have been found to offer superior performance on the majority of natural language understanding (NLU) tasks in the past (e.g., ~\cite{zhao_survey_2023}), open LLMs offer the benefit of being operational locally, thus allowing for better reproducibility of results and supporting technological sovereignty and democratization. 
Therefore, we track the progress of both categories of LLMs over time, to investigate whether open models are catching up with their closed counterparts.

\subsection{Models} \label{sec:models}
The goal of our study was to compare a broad range of current frontier models, i.e. models that perform exceptionally well on code-related tasks. To investigate the trade-off between cost and performance, we included both, closed or proprietary models and open-source/open-weight models. Model selection was based on recent code-focused leaderboards (Aider's polyglot leaderboard\footnote{https://aider.chat/docs/leaderboards/}, BigCodeBench\footnote{https://huggingface.co/spaces/bigcode/bigcodebench-leaderboard}, and RepairBench\footnote{https://repairbench.github.io/}). In total, we evaluated 13 code langugage models (CLMs).

Of the closed models, we included OpenAI's \emph{GPT-4o}, a high-performance model, and \emph{o3-mini}, a lightweight yet capable variant. From Anthropic, we selected \emph{Claude 3.7 Sonnet} for its strong reasoning abilities, along with the smaller \emph{Claude 3.5 Haiku}. Google's \emph{Gemini 1.5 Pro} and the speed-optimised \emph{Gemini 2.0 Flash} concluded the set of proprietary models.   

For the open models, we included \emph{CodeLlama 13B} and \emph{70B} from Meta, both specifically trained for coding tasks. We further evaluated \emph{Llama 3.3}, a more recent multi-purpose LLM with competitive performance. From DeepSeek, we included \emph{DeepSeek Coder 33B} and \emph{DeepSeek R1} in a distilled variant aligned with Llama 70B. We also evaluated Alibaba's \emph{Qwen 2.5 Coder 33B} for its multilingual coding strengths, and \emph{CodeGemma 7B}, Google's lightweight open coding model.

\subsection{Benchmarks} \label{sec:benchmarks}
Our cross-language study uses defect datasets in four widely used programming languages, selected based on the following criteria:
First, the benchmarks should cover a diverse range of common programming languages. Second, they must consist of real bug fixes in real-world programs, excluding artificial data, toy programs, or competition entries. Third, we require code samples in the benchmarks to be executable, i.e. include a test suite with at least one relevant test case per bug, exposing the faulty behavior. Fourth, we only use benchmarks that provide a ground-truth patch written by a human developer. 

Finally, the benchmarks should offer a sufficient number of bugs that are still reproducible today. As Zhu et al.~\cite{DBLP:conf/icse/ZhuR23} have shown, reproducibility is often a challenge in executable software defects datasets, as legacy versions of software artifacts are prone to breakage due to missing/deprecated dependencies or external resources.

Similar to prior studies~\cite{alpharepair,Xia2023,Ouyang2024}, we identified bugs of comparable complexity within the benchmarks to enable fair comparisons across programming languages. To do so, we evaluated single-{\it function} fixes within the benchmarks, meaning we considered only those bugs that could be resolved by modifications within a single function. Unlike many previous studies, however, we did not restrict our analysis to single-{\it line}~\cite{tbar, kolak2022patch} or single-{\it hunk} fixes~\cite{alpharepair}, that is, bugs that can be resolved by modifying a single contiguous piece of code.

Based on these criteria, we selected the following benchmarks in Java, Python, JavaScript, and PHP:


\paragraph{\defectsFourJ v3.0~\cite{DBLP:conf/issta/JustJE14}}
 This curated collection of reproducible bugs, supported by a robust infrastructure, has been widely adopted in software engineering research. Most prior studies on LLM-based APR were performed on \defectsFourJ~\cite{DBLP:conf/issta/JustJE14}, thus enabling a direct comparison with prior work. It contains 854 bugs from 17 open-source Java projects. For our first research question, we used the 264 single-hunk bugs that could be reproduced. 
\paragraph{BugsInPy~\cite{DBLP:conf/sigsoft/WidyasariSLQPTT20}} 
This benchmark dataset comprises 493 real bugs from 17 open-source Python projects across diverse domains, including machine learning, developer tools, scientific computing, and web frameworks. Unlike \defectsFourJ, which distinguishes between active and deprecated bugs (i.e. those no longer reproducible with current Java versions), BugsInPy does not track the reproducibility status of its bugs. Aguilar et al.~\cite{DBLP:conf/scam/AguilarGM23} evaluated the reproducibility of the dataset three years after its initial release and found that only $67\%$ of the expected results could be reproduced in the unmodified BugsInPy dataset. Of these 337 reproducible bugs, 82 were single-hunk bugs, which we used in our first experiment.

\paragraph{BugsJS~\cite{bugsjs2023}}
BugsJS is a collection of JavaScript bugs, collected from 10 widely used server-side JavaScript projects based on Node.js. BugsJS features 453 actual bugs that have been verified manually. 
Similar to BugsInPy, many of the BugsJS subjects suffer from reproducibility issues. Out of the original 139 single-hunk bugs, we identified 72 reproducible cases with at least one relevant test case.

\paragraph{BugsPHP~\cite{DBLP:conf/msr/PramodSTSW24}} BugsPHP is a benchmark dataset of PHP bugs in real-world applications consisting of a training and test dataset. We use BugsPHP's test split, which contains 513 bug-fixing commits from 15 popular open-source PHP projects covering the time period from Jan 2020 to Mar 2023. Each commit is equipped with at least one relevant failing test case. For our initial experiment, we identified 129 reproducible single-hunk bugs.

\begin{table}[]
\caption{Fraction of single-file, single-function, single-hunk and single-line bugs in the benchmarks.}
\begin{tabular}{@{}lrrrrr@{}}
\toprule
Benchmark & all bugs & single-file& single-function& single-hunk& single-line\\ \midrule
\defectsFourJ & 845      & 703         & 471             & 276         & 151         \\
BugsInPy  & 493      & 280         & 189             & 132         & 52          \\
BugsJS    & 513      & 272         & 207             & 139         & 69          \\
BugsPHP   & 452      & 406         & 282             & 178         & 80          \\ \bottomrule
\end{tabular}
\label{tab:complexities}
\end{table}

\subsection{Evaluation Protocol} \label{sec:metrics}

For a meaningful comparison between models, we select metrics that capture key aspects of APR. The goal is to generate patches that fix a bug without introducing new issues. A valid patch must be syntactically correct, parseable, and compilable, as well as functionally correct, meaning it resolves the original issue while preserving existing functionality. 
When it comes to evaluating these criteria, two issues arise: 

\paragraph{Test-Overfitting in APR}  
A challenge in execution-based evaluation of APR is test overfitting~\cite{10.1145/2786805.2786825}: A patch may pass all tests but still fail to implement the intended functionality. To address this issue, the literature distinguishes between \emph{plausible} patches (those that pass the test suite) and \emph{correct} patches (those that satisfy the intended requirements). Since a manual review of patches is infeasible at scale, various proxy metrics, such as Syntactic Equivalence with the developer patch (SYE) and Trivial Compiler Equivalence (TCE), where the patch compiles to equivalent bytecode, have been proposed to approximate patch correctness~\cite{Ouyang2024}. However, while these metrics display some correlation with correct patch quality, they have not outperformed plausibility consistently as predictors. Ouyang et al.~\cite{Ouyang2024} examined the relationship between manually verified correct patches and less expensive metrics (namely, plausibility, TCE, and SYE), and found that TCE and plausibility strongly correlate with patch correctness. A more recent study~\cite{10.1145/3663529.3663776} analyzed test overfitting across several APR approaches and concluded that overfitting may be less problematic than previously assumed. 
Based on these findings, we adopt plausibility (i.e., the passing of test cases) as our primary evaluation criterion.

\paragraph{Non-determinism of LLMs} Due to the inherent non-determinism of LLM-generated output, results may vary across runs, posing a challenge for achieving reliable, reproducible results~\cite{DBLP:journals/corr/abs-2308-02828, DBLP:conf/icse/SallouDP24, 10.1145/3697010}. To mitigate the problem, we sample multiple patches per bug from the same LLM, and report the percentage of bugs for which at least one generated patch passes all tests. More precisely, we adopt the approach by Chen et al.~\cite{DBLP:journals/corr/abs-2107-03374}, generating $n \geq k$ samples per task and computing an unbiased estimator of $pass@k$ as:
\begin{equation}
    pass@k := \underset{Problems}{\mathbb{E}}\left[ 1-\frac{\binom{n-c}{k}}{\binom{n}{k}}\right] 
\end{equation}
where $c$ is the number of plausible samples among $n$ generations. We report \textit{pass@k} for $k=1,5$. 
Based on the standard deviation analysis of \textit{pass@1} for LLM-based APR from~\cite{DBLP:journals/corr/abs-2404-05520}, we use $n=15$ as a reasonable, yet manageable number of generations. We distributed the generation across three independent runs, each generating five candidate patches using each respective model's standard setting with a temperature of $1.0$.  

\paragraph{Significance testing} To assess whether observed differences in $pass@k$ between models are statistically meaningful, we apply the Wilcoxon signed-rank test at a significance level of $\alpha=0.05$. Results are considered significant when the null hypothesis of equal performance can be rejected. In tables comparing model performance, the best result per language is shown in bold, and all results that are not significantly different from the best are underlined.




\subsection{Prompt Design}\label{sec:prompt-design}
The goal of this experimental study is to evaluate the automated program repair capabilities of various large language models across different programming languages in a standardized comparison.
For this purpose, we use a consistent prompt design across all models and benchmarks. All experiments follow a zero-shot prompt setup, where the prompt does not include any example solutions to avoid potential bias from example quality. Furthermore, the prompts are not iterative~\cite{DBLP:journals/corr/abs-2403-17134, DBLP:journals/tosem/ZhangFMSC24, DBLP:journals/corr/abs-2407-01489, autocoderover}, meaning that each model processes a sample/query without follow-up interactions. This choice allows us to maintain a comparable scope for the experiments, ensuring that the results reflect the inherent capabilities of the models without being influenced by more complex techniques built on top of them.


We employ four slightly different prompt templates for each of our experiments (for detailed examples, please refer to our supplemental material):

\begin{lstlisting}[caption={Basic prompt template, example for Java.},label={lst:base-prompt}]
System: You are an automatic program repair tool. Your task is to fix the provided buggy code written in java.

User: The following code snippet contains a bug:

```java
   (*\colorbox{gray!10}{\texttt{< FUNCTION CODE >}} *) 
```

Please provide a fixed version of the buggy function, and only that function without additional imports. Make sure the code is between ``` and ```.

\end{lstlisting}

\paragraph{"Base" Prompt}
The first prompt is used to evaluate LLMs' out-of-the-box APR capabilities. It consists of an initial system directive, followed by the code of the buggy method. An example is shown in Listing \ref{lst:base-prompt}.
While previous work has demonstrated that including the failing test case and the resulting error message can improve the bug-fixing performance of LLMs~\cite{chatrepair, DBLP:journals/corr/abs-2404-05520}, we intentionally adhere to this minimal prompt format to simulate real-world APR scenarios where such test information may be unavailable. The specific prompt is largely adopted from prior work~\cite{DBLP:journals/corr/abs-2404-05520}, with no prompt engineering or optimization applied to prevent introducing model-specific bias.

\begin{lstlisting}[caption={Test prompt template, example for Java.},label={lst:test-prompt}]
The code fails the following tests:

Test (*\colorbox{gray!10}{\texttt{<TEST FILE NAME>::<TEST NAME>}} *):
            
```java
 (*\colorbox{gray!10}{\texttt{< TEST CODE >}} *) 
```

The error message is:

```
 (*\colorbox{gray!10}{\texttt{< ERROR MESSAGE >}} *) 
```

\end{lstlisting}

\paragraph{"Test" Prompt}
The second, more informative prompt incorporates additional information from the failed test cases. It extends the base prompt with the block shown in Listing  \ref{lst:test-prompt}. In addition to the buggy function, the prompt includes the names and source code of the failing test cases along with the corresponding error messages. 
Specifically, we adopt the prompt template from RepairBench~\cite{DBLP:journals/corr/abs-2409-18952} to ensure comparability.

\paragraph{"Perfect Line-level Localization" (Perfect FL / LL) Prompt}
To investigate the impact of fault localization (FL) on APR performance, 
we simulated \emph{perfect line-level localization}, i.e. the \emph{base} prompt is enriched with explicit hints that highlight the correct location of the buggy code segments requiring modification. In preliminary experiments, we evaluated different methods for integrating such line-level localization information into APR prompts, drawing from prior work~\cite{silva2024repairllama, Xia2023} as well as developer intuition. The most effective strategy was found to be a simple comment "\texttt{TODO: Fix here:}" placed immediately before each buggy code snippet, as shown in Listing \ref{lst:ll-prompt}. Note that the end of the buggy code segment was not marked. Based on these findings, we adopted this approach for all perfect-FL experiments.

\begin{lstlisting}[caption={Perfect line-level fault localization prompt template, example for Java.},label={lst:ll-prompt}]
```java
   // Program to fix:
    <M extends Map<String, String>> M putIn(final M map) {
        for (final Entry<String, Integer> entry : mapping.entrySet()) {
            // TODO: fix here
            final int col = entry.getValue().intValue();
                map.put(entry.getKey(), values[col]);
        }
        return map;
    } 
```
\end{lstlisting}

\paragraph{"Automated (Function-level) Localization" (Automated FL) Prompt}
Regarding localization, we also investigate a more realistic setup by drawing the candidate locations from an automated fault localization tool, thus avoiding the assumption of perfect fault localization. In our experiments on \defectsFourJ, we employed the spectrum-based tool FLACOCO~\cite{Silva2021FLACOCOFL}. For each defect, the top three suspicious functions identified by FLACOCO were included in the \emph{base} prompt, as illustrated in Listing \ref{lst:flacoco-prompt}. In instances where a candidate location could not be parsed into a valid function, it was replaced by the next highest-ranked valid function. This configuration mirrors practical scenarios where fault localization is imperfect yet still informative.

\begin{lstlisting}[caption={Automated fault localization prompt template, example for Java.}, label={lst:flacoco-prompt}]
System: You are an automatic program repair tool. Your task is to fix the provided buggy code written in java.

User: I have a buggy program that does not pass all the tests. 
Here are some candidate functions that might need fixing: 

```java
   (*\colorbox{gray!10}{\texttt{< FUNCTION CODE >}} *) 
```

```java
     (*\colorbox{gray!10}{\texttt{< FUNCTION CODE >}} *) 
```

```java
     (*\colorbox{gray!10}{\texttt{< FUNCTION CODE >}} *) 
```

 Please identify the buggy function and provide a fixed version. Return only the fixed function without additional imports. Make sure the code is between ``` and ```.

\end{lstlisting}





%% file: results.tex
\section{Results and Analysis}
\input{values_java}
\input{values_js}
\input{values_php}
\input{values_python}
\input{values}

\subsection{RQ1: APR Performance of current LLMs across different programming languages}

Our first research question aims at comparing LLMs' language- and prompt-specific APR capabilities. We first present an overall impression across all $13$ models and four languages (Section \ref{sec:res_RQ1_basic}), then investigate the influence of adding test information to the prompt (Section \ref{sec:res_RQ1_test}), and finally address the question whether LLMs tend to perform well on specific languages  (Section \ref{sec:res_RQ1_languages}). 


%

\begin{table*}[th]
\centering
\caption{pass@1 and pass@5 results using a simple prompt that contains only the buggy function. For each language, the best result is shown in bold, and all results not significantly different from the best according to the Wilcoxon test ($\alpha=0.05$) are underlined.}
\resizebox{\textwidth}{!}{%
\begin{tabular}{@{}llrrlrrlrrlrrlrr@{}}
\toprule
Model                   &     & \multicolumn{2}{r}{Java}                                        && \multicolumn{2}{r}{JavaScript}                               && \multicolumn{2}{r}{PHP}                                        &&\multicolumn{2}{r}{Python} &&\multicolumn{2}{r}{average} \\ \cmidrule(lr){3-4} \cmidrule(lr){6-7} \cmidrule(lr){9-10} \cmidrule(l){12-13}  \cmidrule(l){15-16}
                        & &pass@$1$                         &pass@$5$                            &&pass@$1$                          &pass@$5$                           &&pass@$1$                            &pass@$5$                           &&pass@$1$       &pass@$5$   &&p@$1$ &pass@$5$ \\ \midrule
Claude 3.7 Sonnet       &     &$\mathbf{\claudeJavaallSHBASICPI}$&$\mathbf{\claudeJavaallSHBASICPV}$&& $\underline{\claudeJSallSHBASICPI}$  & $\underline{\claudeJSallSHBASICPV}$       && $\underline{\claudePHPallSHBASICPI}$       & $\claudePHPallSHBASICPV$      && $\underline{\claudePythonallSHBASICPI}$ & $\underline{\claudePythonallSHBASICPV}$ &&$\underline{\claudeCrossLanguageMeanBASICPI}$& $\underline{\claudeCrossLanguageMeanBASICPV}$\\
Claude 3.5 Haiku        &     & $\claudeHaikuJavaallSHBASICPI$ &$ \claudeHaikuJavaallSHBASICPV$ && $\mathbf{\claudeHaikuJSallSHBASICPI}$ & $\mathbf{\claudeHaikuJSallSHBASICPV}$  && $\claudeHaikuPHPallSHBASICPI$  & $\claudeHaikuPHPallSHBASICPV$ && $\underline{\claudeHaikuPythonallSHBASICPI}$& $\underline{\claudeHaikuPythonallSHBASICPV}$ &&$\claudeHaikuCrossLanguageMeanBASICPI$&$\claudeHaikuCrossLanguageMeanBASICPV$   \\
Gemini 1.5 Pro          &     & $\underline{\geminiIproJavaallSHBASICPI}$  & $\underline{\geminiIproJavaallSHBASICPV}$  && $\underline{\geminiIproJSallSHBASICPI}$  &  $\underline{\geminiIproJSallSHBASICPV}$  && $\underline{\geminiIproPHPallSHBASICPI}$   &  $\underline{\geminiIproPHPallSHBASICPV}$ && $\geminiIproPythonallSHBASICPI$ & $\underline{\geminiIproPythonallSHBASICPV}$        &&$\mathbf{\geminiIproCrossLanguageMeanBASICPI}$&$\underline{\geminiIproCrossLanguageMeanBASICPV}$     \\
Gemini 2.0 Flash        &     &$\underline{\geminiIIflashJavaallSHBASICPI}$&$\geminiIIflashJavaallSHBASICPV$&&$\underline{\geminiIIflashJSallSHBASICPI}$& $\underline{\geminiIIflashJSallSHBASICPV}$&& $\underline{\geminiIIflashPHPallSHBASICPI}$&$\geminiIIflashPHPallSHBASICPV$&& $\mathbf{\geminiIIflashPythonallSHBASICPI}$ & $\mathbf{\geminiIIflashPythonallSHBASICPV  }$  &&$\underline{\geminiIIflashCrossLanguageMeanBASICPI} $&$\underline{\geminiIIflashCrossLanguageMeanBASICPV} $   \\
OpenAI o3-mini          &     & $\underline{\oIIIminiJavaallSHBASICPI}$    & $\oIIIminiJavaallSHBASICPV$    && $\oIIIminiJSallSHBASICPI$    &  $\underline{\oIIIminiJSallSHBASICPV}$    && $\oIIIminiPHPallSHBASICPI$   & $\oIIIminiPHPallSHBASICPV$             && $\underline{\oIIIminiPythonallSHBASICPI}$ & $\underline{\oIIIminiPythonallSHBASICPV}$&&$\underline{\oIIIminiCrossLanguageMeanBASICPI}$&$\oIIIminiCrossLanguageMeanBASICPV$   \\
OpenAI GPT-4o           &     & $\gptIVoJavaallSHBASICPI$      & $\gptIVoJavaallSHBASICPV$      && $\underline{\gptIVoJSallSHBASICPI}$      & $\underline{\gptIVoJSallSHBASICPV}$       && $\gptIVoPHPallSHBASICPI$& $\gptIVoPHPallSHBASICPV$ && $\gptIVoPythonallSHBASICPI$ & $\gptIVoPythonallSHBASICPV$ &&$\gptIVoCrossLanguageMeanBASICPI $&$\underline{\gptIVoCrossLanguageMeanBASICPV} $ \\ \cmidrule(lr){3-4} \cmidrule(lr){6-7} \cmidrule(lr){9-10} \cmidrule(lr){12-13} \cmidrule(l){15-16} 
\deepseekROneDist & 70B & $\RIJavaallSHBASICPI$          & $\RIJavaallSHBASICPV$          && $\underline{\RIJSallSHBASICPI}$ & $\underline{\RIJSallSHBASICPV}$                    && $\mathbf{\RIPHPallSHBASICPI}$  & $\mathbf{\RIPHPallSHBASICPV}$ && $\underline{\RIPythonallSHBASICPI}$ & $\underline{\RIPythonallSHBASICPV}$     &&$\underline{\RICrossLanguageMeanBASICPI}$&$\mathbf{\RICrossLanguageMeanBASICPV}$    \\
Meta Llama 3.3          & 70B & $\LlamaIIIJavaallSHBASICPI$    & $\LlamaIIIJavaallSHBASICPV$    && $\LlamaIIIJSallSHBASICPI$ & $\LlamaIIIJSallSHBASICPV$        && $\LlamaIIIPHPallSHBASICPI$ & $\LlamaIIIPHPallSHBASICPV$ && $\LlamaIIIPythonallSHBASICPI$ & $\LlamaIIIPythonallSHBASICPV$ &&$\LlamaIIICrossLanguageMeanBASICPI$&$\LlamaIIICrossLanguageMeanBASICPV$  \\
Meta CodeLlama          & 70B & $\CLlamaLJavaallSHBASICPI$     & $\CLlamaLJavaallSHBASICPV$     && $\cLlamaLJSallSHBASICPI$ & $\cLlamaLJSallSHBASICPV$  && $\CLlamaLPHPallSHBASICPI$ & $\CLlamaLPHPallSHBASICPV$ && $\CLlamaLPythonallSHBASICPI$ & $\CLlamaLPythonallSHBASICPV$&&$\CLlamaLCrossLanguageMeanBASICPI$&$\CLlamaLCrossLanguageMeanBASICPV$  \\
DeepSeek Coder          & 32B &  $\DSCoderJavaallSHBASICPI$    &  $\DSCoderJavaallSHBASICPV$    && $\DSCoderJSallSHBASICPI$ & $\DSCoderJSallSHBASICPV$          && $\DSCoderPHPallSHBASICPI$  & $\DSCoderPHPallSHBASICPV$  && $\DSCoderPythonallSHBASICPI$ & $\DSCoderPythonallSHBASICPV$         &&$\DSCoderCrossLanguageMeanBASICPI$&$\DSCoderCrossLanguageMeanBASICPV$     \\
Qwen 2.5 Coder          & 32B & $\qwenJavaallSHBASICPI$        & $\qwenJavaallSHBASICPV$        && $\qwenJSallSHBASICPI$ & $\qwenJSallSHBASICPV$                && $\qwenPHPallSHBASICPI$ & $\qwenPHPallSHBASICPV$ && $\qwenPythonallSHBASICPI$ & $\qwenPythonallSHBASICPI$     &&$\qwenCrossLanguageMeanBASICPI$&$\qwenCrossLanguageMeanBASICPV$        \\
Meta CodeLlama          & 13B & $\cLlamaSJavaallSHBASICPI$     & $\cLlamaSJavaallSHBASICPV$     && $\cLlamaSJSallSHBASICPI$   & $\cLlamaSJSallSHBASICPV$        && $\cLlamaSPHPallSHBASICPI$ & $\cLlamaSPHPallSHBASICPV$  && $\cLlamaSPythonallSHBASICPI$ & $\cLlamaSPythonallSHBASICPV $&&$\cLlamaSCrossLanguageMeanBASICPI$&$\cLlamaSCrossLanguageMeanBASICPV$  \\
Google CodeGemma        & 7B  & $\gemmaJavaallSHBASICPI$       & $\gemmaJavaallSHBASICPV$       && $\gemmaJSallSHBASICPI$     &  $\gemmaJSallSHBASICPV$         && $\gemmaPHPallSHBASICPI$ & $\gemmaPHPallSHBASICPV$ && $\gemmaPythonallSHBASICPI$ & $\gemmaPythonallSHBASICPV$ &&$\gemmaCrossLanguageMeanBASICPI$&$\gemmaCrossLanguageMeanBASICPV$ \\ \bottomrule      
\end{tabular}
}
\label{tab:all_models}
\end{table*}

\subsubsection{Overall Results with Base Prompt}
\label{sec:res_RQ1_basic}
\label{sec:res_RQ1}

Table~\ref{tab:all_models} summarizes the performance of all models across the four benchmarks, reporting \emph{pass@1} (the success rate when generating a single fix) and \emph{pass@5} (success among five generated candidates). 
Models were evaluated on all reproducible single-hunk bugs using the basic prompt, which contained only the buggy function. 
The upper section of the table lists closed models, while the lower section presents open models. 
Most models achieve the best performance on \defectsFourJ, only for \deepseekROneDist we see a slightly better performance on BugsPHP with $+0.86\%$ pass@1 and $+2.08\%$ pass@5. However, performance varies across languages, and no model consistently outperforms others across all benchmarks.
For Java, Claude 3.7 Sonnet performs best with $\claudeJavaallSHBASICPI$ pass@1, and $\claudeJavaallSHBASICPV$ pass@5. For Javascript, best results are achieved by Claude 3.5 Haiku with $\claudeHaikuJSallSHBASICPI$ pass@1 and $\claudeHaikuJSallSHBASICPV$ pass@5.
On BugsPHP, the distilled version of DeepSeek R1 (distilled to Llama 3 70B) achieves the best pass@1 at $\RIPHPallSHBASICPI$ and pass@5 at $\RIPHPallSHBASICPV$, %
while for Python Gemini 2.0 Flash shows the best performance, reaching $\geminiIIflashPythonallSHBASICPI$ pass@1 and $\geminiIIflashPythonallSHBASICPV$ pass@5.

\begin{figure}[h]
    \centering
    \includegraphics[width=0.7\linewidth]{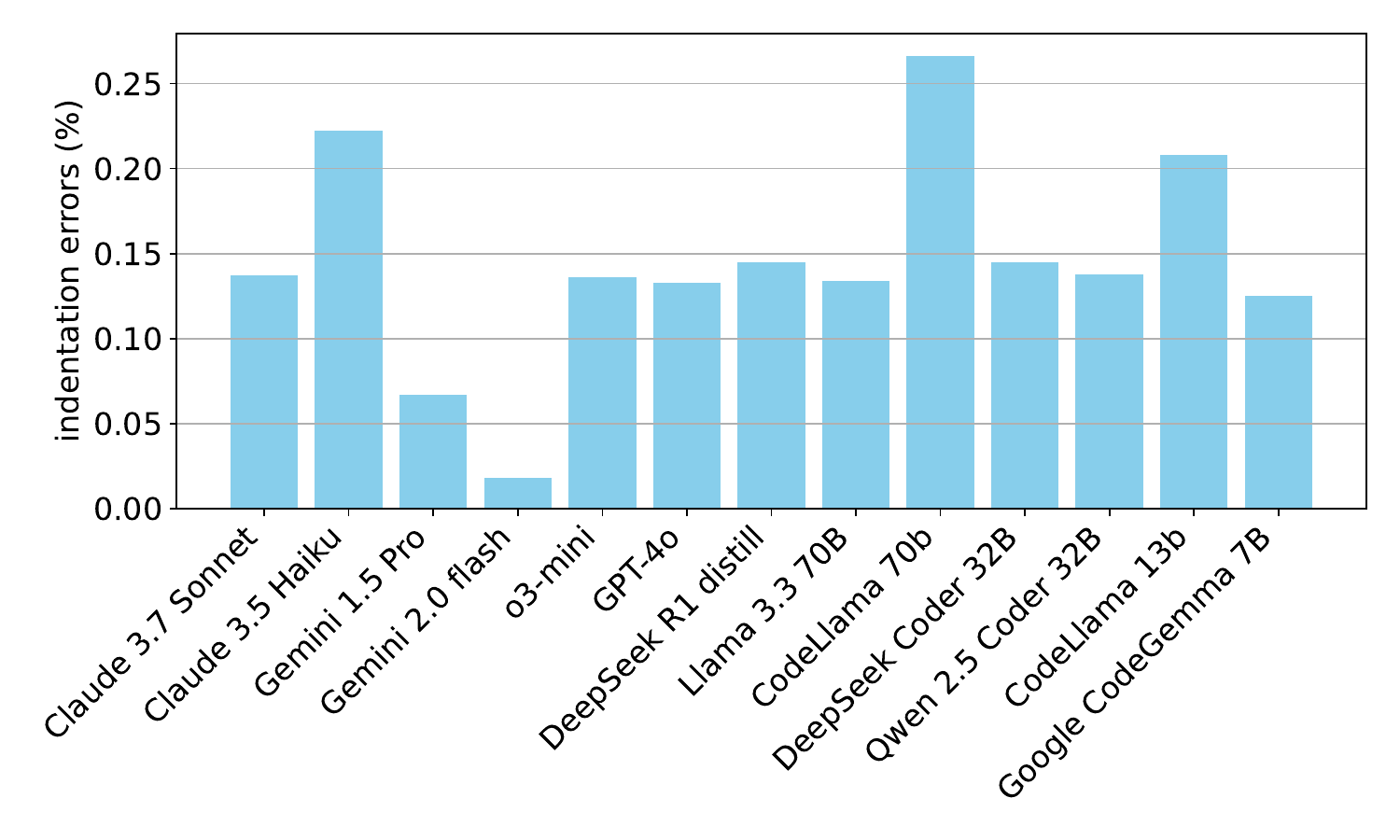}
    \caption{Frequency of indentation errors for the Python benchmark. Notably, most models struggle with indentation issues, with the Gemini models being the exception.}
    \label{fig:python_indentation}
\end{figure}

A notable observation is the surprisingly poor performance of all models on the Python benchmark. Further examination of the generated patches reveals that this is largely due to indentation issues in the generated patches. While other programming languages are more forgiving, Python’s off-side syntax rule leads to execution failures when code is misaligned.
Figure~\ref{fig:python_indentation} illustrates the percentage of indentation errors in the generated patches. Notably, the Gemini models, particularly Gemini 2.0 Flash, produce significantly fewer indentation errors than other models. This is reflected in the comparatively good performance on the Python benchmark. 
This finding suggests that when applying APR with LLMs to Python, practitioners should consider integrating a linting or formatting step into their post-processing pipeline to mitigate syntax-related issues.

\subsubsection{Influence of Test Information}
\label{sec:res_RQ1_test}
Next, we compare the {\it base} prompt, which contains only the faulty method and a generic repair instruction, with the {\it test} prompt, which extends the base prompt to include the names, source code, and error messages of failing test cases.
For this and all subsequent experiments, we broadened the evaluation scope from single-hunk to single-function bugs. 
To keep the dataset size manageable, we selected a representative subset of 100 bugs per benchmark using stratified sampling over the three complexity levels: We first analyzed the distribution of single-line, single-hunk, and multi-hunk patches per benchmark (see Table~\ref{tab:complexities}) and sampled accordingly to preserve each benchmark's original complexity profile. Additionally, we ensured that all repositories included in the original benchmarks were represented in the sampled data. The final distribution is summarized in Table~\ref{tab:complexities_selection}. 

Table \ref{tab:prompts} includes a comparison of both prompts for the most promising models from Table \ref{tab:all_models}. These include the four best-performing closed-source models (averaged across all languages) and, among the open models, the best large model and the best medium-sized model. 
Please note that the figures for the base prompt differ from Table~\ref{tab:all_models}, as the evaluation data is different. We observe that adding test information (rows \emph{Test}) yields substantial improvements compared to the base prompt, reaching up to almost $+47\%$ pass@1 (\deepseekROneDist on Python). This confirms results of prior work~\cite{chatrepair, DBLP:journals/corr/abs-2404-05520} for the most recent models, and also across multiple programming languages. While all models pick up on the additional test information, the effect appears to be stronger for certain models (particularly for OpenAI's o3-mini). Notably, the improvements by test information show less variation between the different {\it languages}: Across all four benchmarks, test information consistently improves results, with Python ($+34.7$\% pass@1, averaged over all models) showing the strongest improvements.



\begin{table}[ht]
\centering
\caption{Distribution of multi-hunk, single-hunk and single-line bugs in the evaluation subset. Categories are exclusive.}
\begin{tabularx}{0.9\columnwidth}{@{}Xrrrr@{}}
\toprule
Benchmark & all & multi-hunk &single-hunk& single-line\\ \midrule
\defectsFourJ      & 100         & 41             & 26         & 33         \\
BugsInPy      & 100         & 30             & 42         & 28          \\
BugsJS         & 100         & 32             & 34         & 34         \\
BugsPHP        & 100         & 36             & 35         & 29          \\ \bottomrule
\end{tabularx}
\label{tab:complexities_selection}
\end{table}


\begin{table}[ht]
\caption{$Pass@1$ and $pass@5$ with varying prompt ingredients: the \textit{Base} prompt includes only the buggy function; \textit{Test} includes failing test case information; \textit{LL} adds line-level fault localization via comments.}
\centering
\resizebox{\textwidth}{!}{
\begin{tabular}{llrrlrrlrrlrr}
\toprule
                  &        & \multicolumn{2}{r}{Java}                                        &  & \multicolumn{2}{r}{JavaScript}                                     &  & \multicolumn{2}{r}{PHP}         &  & \multicolumn{2}{r}{Python} \\ \cmidrule{3-4} \cmidrule{6-7} \cmidrule{9-10} \cmidrule{12-13} 
                  & Prompt & pass@1                          & pass@5                        &  & pass@1                         & pass@5                            &  & pass@1             & pass@5     &  & pass@1       & pass@5      \\ \cmidrule{1-4} \cmidrule{6-7} \cmidrule{9-10} \cmidrule{12-13} 
Claude 3.7 Sonnet & Base   & $\claudeJavaBASICPI$            & $\claudeJavaBASICPV$          &  & $\claudeJSBASICPI$             & $\claudeJSBASICPV$                & & $\claudePHPBASICPI$ & $\claudePHPBASICPV$    && $\claudePythonBASICPI$       & $\claudePythonBASICPV$     \\

                  & Test   & $\claudeJavaTestPI$    & $\claudeJavaTestPV$ &          & $\claudeJSTestPI$   & $\claudeJSTestPV$             &  & $\claudePHPTestPI$   & $\claudePHPTestPV$    &          & $\claudePythonTestPI$     & $\claudePythonTestPV$  \\
                  & LL     & $\claudeJavaPerfectLLPI$  & $\claudeJavaPerfectLLPV$ &      & $\claudeJSPerfectLLPI$      & $\claudeJSPerfectLLPV$    &  & $\claudePHPPerfectLLPI$    & $\claudePHPPerfectLLPV$ &  & $\claudePythonPerfectLLPI$    & $\claudePythonPerfectLLPV$    \\\cmidrule{3-4} \cmidrule{6-7} \cmidrule{9-10} \cmidrule{12-13} 
Gemini 1.5 Pro    & Base   & $\geminiIproJavaBASICPI$        & $\geminiIproJavaBASICPV$     && $\geminiIproJSBASICPI$     & $\geminiIproJSBASICPV$     && $\geminiIproPHPBASICPI$     & $\geminiIproPHPBASICPV$     && $\geminiIproPythonBASICPI$     & $\geminiIproPythonBASICPV$     \\ 
                   
                   & Test & $\geminiIproJavaTestPI$      & $\geminiIproJavaTestPV$      && $\geminiIproJSTestPI$      & $\geminiIproJSTestPV$      && $\geminiIproPHPTestPI$      & $\geminiIproPHPTestPV$      && $\geminiIproPythonTestPI$      & $\geminiIproPythonTestPV$  \\
                   & LL   & $\geminiIproJavaPerfectLLPI$ & $\geminiIproJavaPerfectLLPV$ && $\geminiIproJSPerfectLLPI$ & $\geminiIproJSPerfectLLPV$ && $\geminiIproPHPPerfectLLPI$ & $\geminiIproPHPPerfectLLPV$ && $\geminiIproPythonPerfectLLPI$ & $\geminiIproPythonPerfectLLPV$ \\ \cmidrule{3-4} \cmidrule{6-7} \cmidrule{9-10} \cmidrule{12-13} 
Gemini 2.0 Flash   & Base & $\geminiIIflashJavaBASICPI$   & $\geminiIIflashJavaBASICPV$     && $\geminiIIflashJSBASICPI$     & $\geminiIIflashJSBASICPV$     && $\geminiIIflashPHPBASICPI$     & $\geminiIIflashPHPBASICPV$     && $\geminiIIflashPythonBASICPI$     & $\geminiIIflashPythonBASICPV$     \\ 
                   
                   & Test & $\geminiIIflashJavaTestPI$      & $\geminiIIflashJavaTestPV$      && $\geminiIIflashJSTestPI$      & $\geminiIIflashJSTestPV$      && $\geminiIIflashPHPTestPI$      & $\geminiIIflashPHPTestPV$      && $\geminiIIflashPythonTestPI$      & $\geminiIIflashPythonTestPV$  \\ 
                   & LL   & $\geminiIIflashJavaPerfectLLPI$ & $\geminiIIflashJavaPerfectLLPV$ && $\geminiIIflashJSPerfectLLPI$ & $\geminiIIflashJSPerfectLLPV$ && $\geminiIIflashPHPPerfectLLPI$ & $\geminiIIflashPHPPerfectLLPV$ && $\geminiIIflashPythonPerfectLLPI$ & $\geminiIIflashPythonPerfectLLPV$ \\ \cmidrule{3-4} \cmidrule{6-7} \cmidrule{9-10} \cmidrule{12-13} 
o3-mini            & Base & $\oIIIminiJavaBASICPI$        & $\oIIIminiJavaBASICPV$     && $\oIIIminiJSBASICPI$     & $\oIIIminiJSBASICPV$     && $\oIIIminiPHPBASICPI$     & $\oIIIminiPHPBASICPV$     && $\oIIIminiPythonBASICPI$     & $\oIIIminiPythonBASICPV$     \\ 
                   
                   & Test & $\oIIIminiJavaTestPI$      & $\oIIIminiJavaTestPV$      && $\oIIIminiJSTestPI$      & $\oIIIminiJSTestPV$      && $\oIIIminiPHPTestPI$      & $\oIIIminiPHPTestPV$      && $\oIIIminiPythonTestPI$      & $\oIIIminiPythonTestPV$      \\ 
                   & LL   & $\oIIIminiJavaPerfectLLPI$ & $\oIIIminiJavaPerfectLLPV$ && $\oIIIminiJSPerfectLLPI$ & $\oIIIminiJSPerfectLLPV$ && $\oIIIminiPHPPerfectLLPI$ & $\oIIIminiPHPPerfectLLPV$ && $\oIIIminiPythonPerfectLLPI$ & $\oIIIminiPythonPerfectLLPV$ \\\cmidrule{3-4} \cmidrule{6-7} \cmidrule{9-10} \cmidrule{12-13} 
DeepSeek R1 dist.  & Base &$\RIJavaBASICPI$   &  $\RIJavaBASICPV$ &          & $\RIJSBASICPI$    & $\RIJSBASICPV$ &            & $\RIPHPBASICPI$     & $\RIPHPBASICPV$ &              & $\RIPythonBASICPI$   & $\RIPythonBASICPV$        \\
                  
                   & Test & $\RIJavaTestPI$   &  $\RIJavaTestPV$ &          & $\RIJSTestPI$    & $\RIJSTestPV$ &            & $\RIPHPTestPI$     & $\RIPHPTestPV$ &              & $\RIPythonTestPI$   & $\RIPythonTestPV$        \\
                    & LL   & $\RIJavaPerfectLLPI$   &  $\RIJavaPerfectLLPV$ &   & $\RIJSPerfectLLPI$    & $\RIJSPerfectLLPV$ &            & $\RIPHPPerfectLLPI$     & $\RIPHPPerfectLLPV$ &     & $\RIPythonPerfectLLPI$   & $\RIPythonPerfectLLPV$     \\ \cmidrule{3-4} \cmidrule{6-7} \cmidrule{9-10} \cmidrule{12-13} 
Qwen 2.5 Coder     & Base & $\qwenJavaBASICPI$     & $\qwenJavaBASICPV$     && $\qwenJSBASICPI$     & $\qwenJSBASICPV$     && $\qwenPHPBASICPI$     & $\qwenPHPBASICPV$     && $\qwenPythonBASICPI$     & $\qwenPythonBASICPV$     \\ 
                   
                   & Test & $\qwenJavaTestPI$      & $\qwenJavaTestPV$      && $\qwenJSTestPI$      & $\qwenJSTestPV$      && $\qwenPHPTestPI$      & $\qwenPHPTestPV$      && $\qwenPythonTestPI$      & $\qwenPythonTestPV$      \\ 
                   & LL   & $\qwenJavaPerfectLLPI$ & $\qwenJavaPerfectLLPV$ && $\qwenJSPerfectLLPI$ & $\qwenJSPerfectLLPV$ && $\qwenPHPPerfectLLPI$ & $\qwenPHPPerfectLLPV$ && $\qwenPythonPerfectLLPI$ & $\qwenPythonPerfectLLPV$ \\ \bottomrule

\end{tabular}
 }
\label{tab:prompts}
\end{table}

\begin{figure}[ht]
    \centering
    \includegraphics[width=1\linewidth]{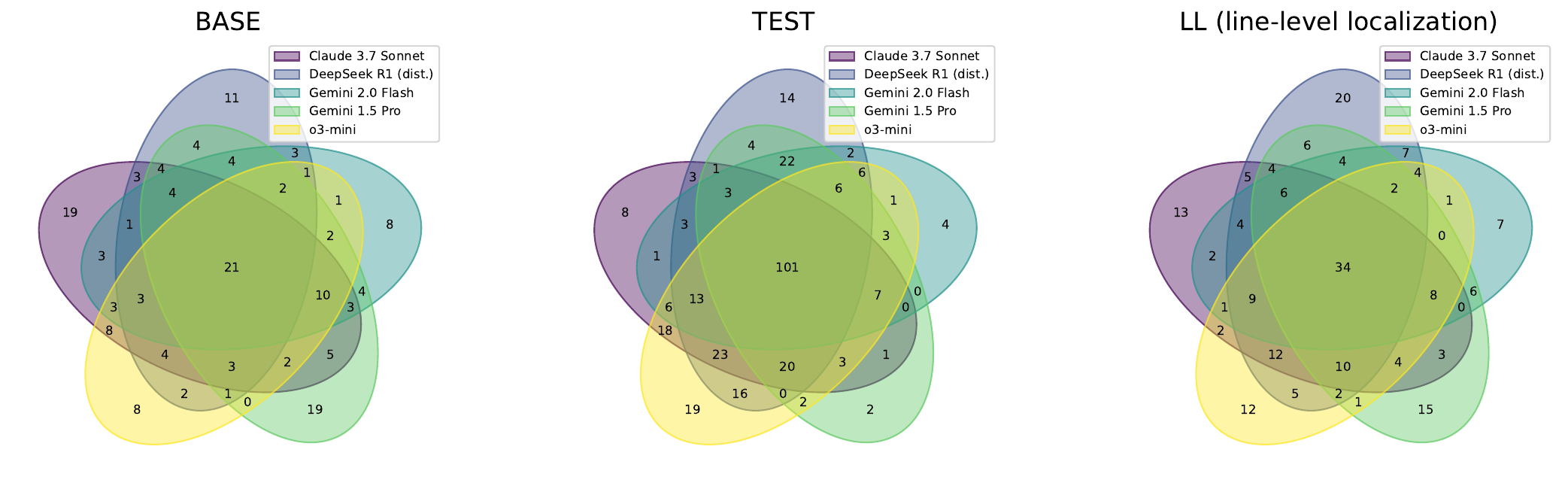}
    \caption{Bug fixes per model, determined for 400 single-function bugs in four programming languages. }
    \label{fig:bug_fix_venn}
    \Description{Venn-diagrams of the bug-fixed per model.}
\end{figure}

\subsubsection{Language Specialization}
\label{sec:res_RQ1_languages}

\begin{figure}[ht]
    \centering
    \includegraphics[width=1\linewidth]{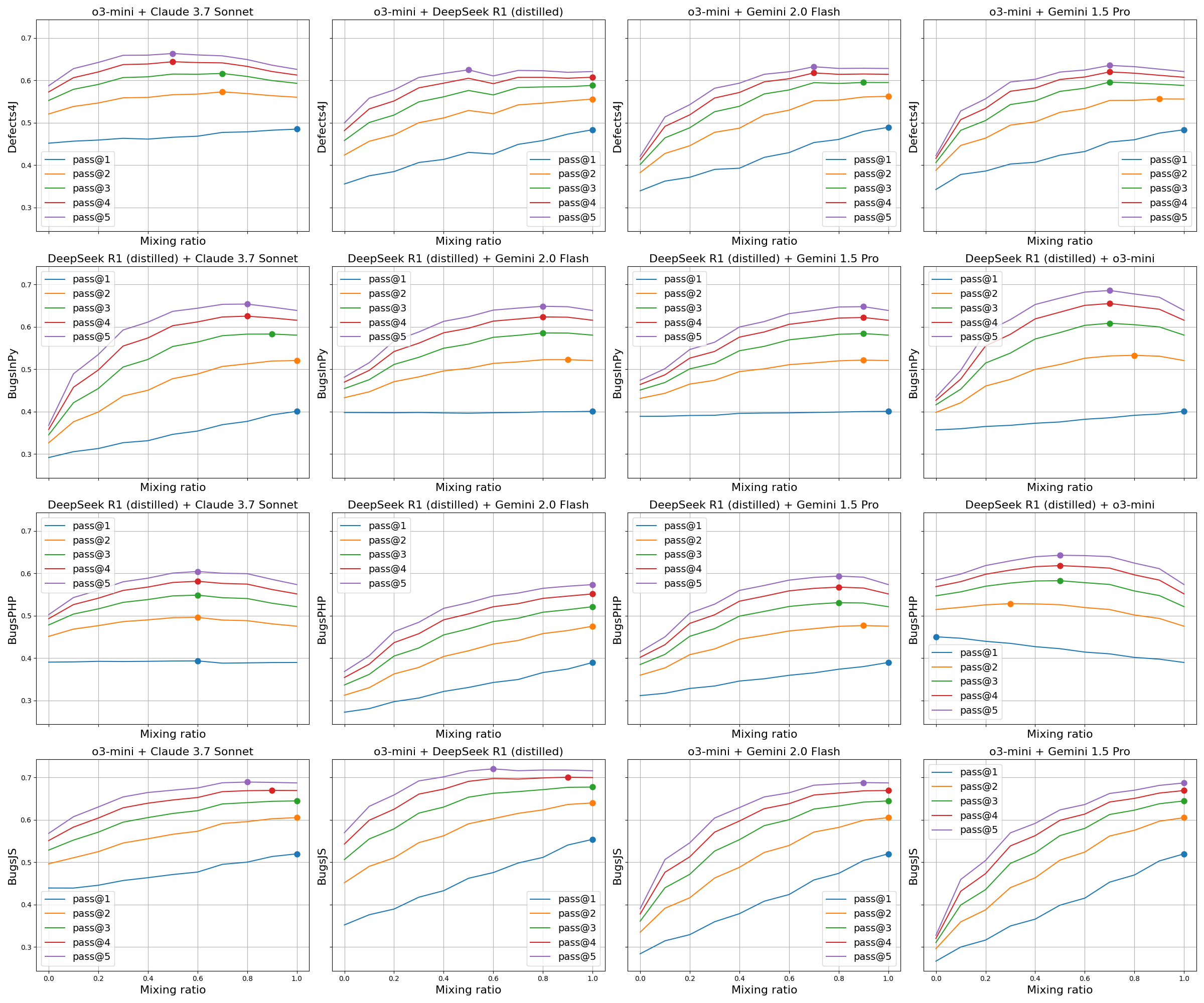}
    \caption{Combined $pass@k$ (using the \emph{test} prompt) for ensembles of the best performing model per benchmark with a second model.}
    \Description{Charts showing combined pass@k results for model ensembles.}
    \label{fig:model_combinations}
\end{figure}

Table \ref{tab:all_models} has already suggested that model accuracy varies between programming languages, with four different LLMs performing best on the four language-specific benchmarks. Since model complementarity and language specialization are important issues for practitioners, we investigate these issues further. 
First, the Venn diagrams in Figure \ref{fig:bug_fix_venn} 
displays what model combinations perform well on which subsets of cases. Here, all cases from all benchmarks were joined, and a model is defined to "solve" an issue if at least one of its $n=15$ generations is plausible.
For each model, we find cases that can only be fixed by this particular model. This effect holds for all three prompt types "basic", "test", and "line-level localization (LL)" (see Section \ref{sec:prompt-design}). 
To further investigate how different models complement each other, we evaluate, how $pass@k$ changes when the $k$ candidates are derived from two models rather than just one. To this end, we calculate $p_{\text{combined}}@k$ when we draw a fraction of $\alpha \in [0,1]$ of the $n=15$ generations from one model and the remaining $(1{-}\alpha) \cdot n$ generations from a second model:
\begin{equation}
    p_{\text{combined}}@k := \underset{\substack{\text{Problems,} \\ \text{Simulations}}}{\mathbb{E}}\left[ 1 - 
    \frac{\binom{n - c_{\text{combined}}}{k}}{\binom{n}{k}} 
    \right] 
\end{equation}
where $ c_{\text{combined}}$ represents the number of correct patches among the $n$ combined generations. To avoid rounding errors, we approximate this process by simulation: for each $\alpha$, we repeatedly (100 times in our experiments) sample $\alpha \cdot n$ generations from the first model and $(1{-}\alpha) \cdot n$ generations from the second model, and determine $c_{\text{combined}}$ as the number of plausible patches in the combined set. We then estimate $p_{\text{combined}}@k$ as the expectation over all simulation runs.  
Figure \ref{fig:model_combinations} shows $p_{\text{combined}}@k$ for ensembles that pair the best-performing model in each benchmark with one of the next four best models, evaluated across different mixing fractions $\alpha$, using the \emph{test} prompt. In most cases, combining two models improved performance over the best individual model, with more pronounced gains at higher values of $k$. 

For instance, with the \emph{test} prompt $pass@5$ improved in 14 out of 16 experiments. Note that, since $\alpha$ corresponds to the models' mixing ratio, the rightmost point in each plot corresponds to sampling only from the best model, and the leftmost point to sampling only from the second model.
 We found $pass@5$ to improve for 
(1) JavaScript from $\oIIIminiJSTestPV$ (o3-mini) to $\oIIIminiRIJSTestPV$ (o3-mini \& \deepseekROneDist), 
(2) Java       from $\oIIIminiJavaTestPV$ (o3-mini) to $\oIIIminiclaudeJavaTestPV$ (o3-mini \& Claude 3.7),         
(3) PHP        from $\oIIIminiPHPTestPV$ (o3-mini) to $\oIIIminiRIPHPTestPV$  (o3-mini \& \deepseekROneDist), and
(4) Python     from  $\RIPythonTestPV$ (\deepseekROneDist) to $\RIclaudePythonTestPV$ (Claude 3.7 \& \deepseekROneDist).

Table \ref{tab:commitees} summarizes the results, reporting for each benchmark and prompt the best single model and the best two-model ensemble, constructed by combining two-thirds of the generations from the top-performing model with one-third from a complementary model.
Overall, the results suggest there are substantial differences in LLM repair performance between programming languages, and demonstrate that LLMs can effectively complement each other on repair tasks. 


\begin{table}[t]
\centering
\caption{Comparison between the strongest standalone models and most effective model combinations ($\frac{2}{3}$ of the generations from the top model and $\frac{1}{3}$ from a complementary model)
 }
\begin{tabular}{llllrllr}
\toprule
           & Prompt &  & Best single model         & pass@5                      && Best ensemble                   & pass@5 \\ \midrule 
Java       & Base   &  & Gemini 1.5 Pro            & $\geminiIproJavaBASICPV$ &&  Gemini 1.5 Pro \& Claude 3.7   & $\claudegeminiIproJavaBASICPV$    \\       
           & Test   &  & o3-mini                   & $\oIIIminiJavaTestPV$    && o3-mini  \& Claude 3.7          & $\oIIIminiclaudeJavaTestPV$    \\
           & LL     &  & Claude 3.7                & $\claudeJavaPerfectLLPV$ && Claude 3.7 \& \deepseekROneDist & $\claudeRIJavaPerfectLLPV$    \\ \cmidrule{1-2} \cmidrule{4-5} \cmidrule{7-8} 
JavaScript & Base   &  & \deepseekROneDist         & $\RIJSBASICPV$           && \deepseekROneDist \& Claude 3.7 & \RIclaudeJSBASICPV    \\
           & Test   &  & o3-mini                   & $\oIIIminiJSTestPV$      && o3-mini \& \deepseekROneDist    &  $\oIIIminiRIJSTestPV$   \\ 
           & LL     &  & Claude 3.7                & $\claudeJSBASICPV$       && Claude 3.7 \& \deepseekROneDist & \claudeRIJSPerfectLLPV   \\ \cmidrule{1-2} \cmidrule{4-5} \cmidrule{7-8} 
Python     & Base   &  & Gemini 2.0                & $\geminiIIflashPythonBASICPV$ && Gemini 2.0 \& Claude 3.7   & \geminiIIflashclaudePythonBASICPV    \\  
           & Test   &  & \deepseekROneDist         & $\RIPythonTestPV$        && Claude 3.7 \& \deepseekROneDist & $\RIclaudePythonTestPV$     \\
           & LL     &  & Gemini 2.0                & $\geminiIIflashPythonPerfectLLPV$ && Gemini 2.0 \& o3-mini  & $\geminiIIflashoIIIminiPythonPerfectLLPV$ \\ \cmidrule{1-2} \cmidrule{4-5} \cmidrule{7-8} 
PHP        & Base   &  & Gemini 1.5 Pro            & $\geminiIproPHPBASICPV$  && Gemini 1.5 Pro \& \deepseekROneDist & $\geminiIproRIPHPBASICPV$    \\       
           & Test   &  & o3-mini                   & $\oIIIminiPHPTestPV$     && o3-mini \& \deepseekROneDist    & $\oIIIminiRIPHPTestPV$    \\ 
           & LL     &  & \deepseekROneDist & $\RIPHPPerfectLLPV$              && \deepseekROneDist \& Gemini 1.5 Pro & $\RIgeminiIproPHPPerfectLLPV$    \\ \bottomrule
\end{tabular}
\label{tab:commitees}
\end{table}

\subsection{RQ2: Impact of Fault Localization on LLM-based APR }

Current work in APR lacks a standardized approach regarding assumptions about fault localization~\cite{renzullo2025automated}. While end-to-end repair --including fault localization (FL), patch generation, and prioritization -- is inherently more challenging than patch generation in isolation, most empirical evaluations of neural program repair assume \emph{perfect fault localization} as suggested by Liu et al.~\cite{DBLP:conf/icst/LiuKB0KT19}, usually derived directly from developer fixes. This assumption enables objective evaluation but also introduces a disconnect from realistic use cases where fault localization is noisy and imprecise. The impact of this assumption remains largely underexplored, raising concerns about localization bias~\cite{DBLP:conf/icsm/YangQM17, DBLP:conf/icst/LiuKB0KT19}, which may lead to an overestimation of model performance under idealized conditions. Therefore, we investigate two aspects: the impact of FL granularity and the effect of noisy, tool-generated FL on patch generation. 

\subsubsection{Granularity of Fault Localization}
We compare the base prompt -- which includes (only) the faulty method and thus simulates perfect method-level FL -- with the Line-level Localization (LL) prompt, which includes a hint about the particular code lines to fix. Table \ref{tab:prompts} displays results for both prompts across the most promising models from RQ1. We see that line-level localization -- while improving results -- appears to add less value compared to test errors for (almost) all languages-model combinations. Also, we observe that the improvements with line-level localization appear to be inconsistent: Particularly, in the PHP benchmark, the accuracy deteriorates for $4$ out of $6$ LLMs when introducing line-level localization. This suggests that future research on localization in APR should evaluate across multiple programming languages.

\subsubsection{Automated Fault Localization}

Assuming perfect (method- or line-level) localization may be a questionable assumption in practice, where the scope of bugs may have to be narrowed down within large software projects~\cite{DBLP:conf/icst/LiuKB0KT19}. To study the relationship between fault localization and neural patch generation in a realistic setting, we use FLACOCO~\cite{Silva2021FLACOCOFL}, a recent spectrum-based localization tool, to identify the three most suspicious functions as potential locations for the fix. We then prompted the model using these candidate functions. As the tool currently supports only Java, this experiment was conducted on \defectsFourJ (using the subselection of $100$ samples). The results are presented in Table \ref{tab:fl_flacoco}. The first two columns show the pass@1 results  
using perfect fault localization. In comparison, pass@1 using automated fault localization in the third column shows a significant drop. The performance drop ranges from $ -7.83$\% for Gemini 2.0 Flash to $-16.20$\% for Claude 3.7 Sonnet compared to the base prompt. This can be attributed to two main factors. First, the accuracy of the automated fault localization tool is lower than that of perfect fault localization. In our setup, the actual fix location appeared among the top-3 candidates for only $28$ out of $100$ faults. For the remaining cases, the prompt included only irrelevant code locations. We discovered only one such case for which two LLMs produced plausible patches, and even these could be attributed to test overfitting.
Also, even when the correct function was identified, we observed a degradation in performance: on average, pass@k decreased by $5.54$\%, and in $9.8$\% of those cases, no plausible patch could be generated -- despite the base prompt yielding plausible solutions within the 15 generations per model. 
Since automated line-level localization can be expected to be even harder (and due to the mixed results with (perfect) line-level localization in Table \ref{tab:prompts}), we did not explore automated line-level localization further.

Overall, our results for RQ2 suggest that practical, automated localization has a significant impact on APR results. Future APR research might consider including practice-oriented evaluations with localization tools, or even put an even stronger focus on integrated approaches of repair and localization, such as \cite{crashrepair}.

\begin{table}[h]
\centering
\caption{pass@1 results on \defectsFourJ, using perfect function-/line-level and automated (top-3 spectrum-based) Fault Localization.}
\begin{tabularx}{0.7\columnwidth}{@{}Xrrrr@{}}
\toprule
                        & \multicolumn{2}{r}{Perfect FL}                             && Automated FL \\ \cmidrule(l){2-3} \cmidrule{5-5} 
                        & Function                   & Line                          && Function                \\ \midrule
Claude 3.7 Sonnet       & \claudeJavaBASICPI         & \claudeJavaPerfectLLPI        && \claudeJavaFlacocoPI               \\
Gemini 1.5 Pro          &  \geminiIproJavaBASICPI    & \geminiIproJavaPerfectLLPI    && \geminiIproJavaFlacocoPI               \\
Gemini 2.0 Flash        &  \geminiIIflashJavaBASICPI & \geminiIIflashJavaPerfectLLPI && \geminiIIflashJavaFlacocoPI               \\
o3-mini                 &  \oIIIminiJavaBASICPI      & \oIIIminiJavaPerfectLLPI      && \oIIIminiJavaFlacocoPI               \\
\deepseekROneDist       & \RIJavaBASICPI             & \RIJavaPerfectLLPI            && \RIJavaFlacocoPI                             \\
Qwen 2.5 Coder          &  \qwenJavaBASICPI          & \qwenJavaPerfectLLPI          && \qwenJavaFlacocoPI             \\ \bottomrule
\end{tabularx}
\label{tab:fl_flacoco}
\end{table}

\subsection{RQ3: Patch Complexity}

We also test how repair accuracy varies for different levels of patch complexity. To do so, Table \ref{tab:bug_complexities_test} reports pass@1 and pass@5 for all benchmarks (using the \emph{test} prompt) separately for single-line, single-hunk, and multi-hunk issues. While we expect complexity to increase with these categories significantly, the expected performance deterioration was not as high as we would have expected. For $pass@1$, we observed a drop from $\averageSLTestPI$ (single-line) over  $\averageSHTestPI$ (single-hunk) to  $\averageMHTestPI$ (multi-hunk), i.e., LLMs cope still relatively well with increasing patch complexity.
We even observe $9$ of $48$ performance figures to improve from single-line to single-hunk (particularly for Python), and $10$ of $48$ from single-hunk to multi-hunk (particularly for Java). 
Note that the experiment focused on single-function cases. However, these promising results suggest to even investigate multi-hunk, multi-function APR further.

\begin{table*}[t]
\centering
\caption{Comparison of different patch complexities using failing testcases in the prompt. Best results per language and complexity are shown in bold, and results not significantly different from the best (Wilcoxon test, $\alpha=0.05$) are underlined.}
\begin{tabular}{@{}llrrrrrrrrr@{}}
\toprule
          &                  && \multicolumn{2}{r}{Single-Line}                         && \multicolumn{2}{r}{Single-Hunk}                         && \multicolumn{2}{r}{Multi-Hunk} \\ \cmidrule{4-5} \cmidrule{7-8} \cmidrule{10-11} 
          &                  &&pass@$1$                        &pass@$5$                        &&pass@$1$                        &pass@$5$                        &&pass@$1$                        &pass@$5$                         \\ \midrule
\defectsFourJ & Claude 3.7 Sonnet&& $\underline{\claudeJavaSLTestPI}$       & $\underline{\claudeJavaSLTestPV}$      && $\underline{\claudeJavaSHTestPI}$      & $\underline{\claudeJavaSHTestPV}$       && $\underline{\claudeJavaMHTestPI}$       & $\underline {\claudeJavaMHTestPV}$ \\
          & Gemini 1.5 Pro   && $\geminiIproJavaSLTestPI$   & $\geminiIproJavaSLTestPV$   && $\underline{\geminiIproJavaSHTestPI}$   & $\geminiIproJavaSHTestPV$   && $\geminiIproJavaMHTestPI$   & $\geminiIproJavaMHTestPV$ \\
          & Gemini 2.0 Flash && $\underline{\geminiIIflashJavaSLTestPI}$& $\geminiIIflashJavaSLTestPV$&& $\geminiIIflashJavaSHTestPI$& $\geminiIIflashJavaSHTestPV$&&$\underline{\geminiIIflashJavaMHTestPI}$& $\underline{\geminiIIflashJavaMHTestPV}$ \\
          & o3-mini          && $\mathbf{\oIIIminiJavaSLTestPI}$     & $\mathbf{\oIIIminiJavaSLTestPV}$     && $\mathbf{\oIIIminiJavaSHTestPI}$    & $\mathbf{\oIIIminiJavaSHTestPV}$     && $\mathbf{\oIIIminiJavaMHTestPI}$   & $\mathbf{\oIIIminiJavaMHTestPV}$           \\
          & \deepseekROneDist      && $\RIJavaSLTestPI$           & $\underline{\RIJavaSLTestPV}$           && $\RIJavaSHTestPI$           & $\RIJavaSHTestPV$           && $\RIJavaMHTestPI$           & $\RIJavaMHTestPV$         \\
          & Qwen 2.5 Coder   && $\qwenJavaSLTestPI$         & $\qwenJavaSLTestPV$         && $\qwenJavaSHTestPI$         & $\qwenJavaSHTestPV$         && $\qwenJavaMHTestPI$        & $\qwenJavaMHTestPV$          \\ \cmidrule{1-2} \cmidrule{4-5} \cmidrule{7-8} \cmidrule{10-11}
BugsJS    & Claude 3.7 Sonnet&& $\underline{\claudeJSSLTestPI}$         & $\claudeJSSLTestPV$         && $\underline{\claudeJSSHTestPI}$         & $\underline{\claudeJSSHTestPV}$         && $\underline{\claudeJSMHTestPI}$        & $\underline{\claudeJSMHTestPV}$              \\
          & Gemini 1.5 Pro   && $\underline{\geminiIproJSSLTestPI}$     & $\geminiIproJSSLTestPV$     && $\geminiIproJSSHTestPI$     &  $\geminiIproJSSHTestPV$    && $\geminiIproJSMHTestPI$     &  $\geminiIproJSMHTestPV$   \\
          & Gemini 2.0 Flash && $\geminiIIflashJSSLTestPI$  & $\geminiIIflashJSSLTestPV$  && $\geminiIIflashJSSHTestPI$  & $\geminiIIflashJSSHTestPV$  && $\geminiIIflashJSMHTestPI$  & $\geminiIIflashJSMHTestPV$ \\
          & o3-mini          && $\mathbf{\oIIIminiJSSLTestPI}$       & $\mathbf{\oIIIminiJSSLTestPV}$       && $\mathbf{\oIIIminiJSSHTestPI}$      & $\mathbf{\oIIIminiJSSHTestPV}$       && $\mathbf{\oIIIminiJSMHTestPI}$      & $\mathbf{\oIIIminiJSMHTestPV}$ \\
          & \deepseekROneDist && $\RIJSSLTestPI$ & $\underline{\RIJSSLTestPV}$    && $\RIJSSHTestPI$      & $\RIJSSHTestPV$ && $\RIJSMHTestPI$ & $\RIJSMHTestPV$               \\
          & Qwen 2.5 Coder   && $\qwenJSSLTestPI$           & $\qwenJSSLTestPV$           && $\qwenJSSHTestPI$           & $\qwenJSSHTestPV$          && $\qwenJSMHTestPI$           &  $\qwenJSMHTestPV$        \\  \cmidrule{1-2} \cmidrule{4-5} \cmidrule{7-8} \cmidrule{10-11}
BugsInPy & Claude 3.7 Sonnet && $\claudePythonSLTestPI$     & $\claudePythonSLTestPV$     && $\underline{\claudePythonSHTestPI}$     & $\claudePythonSHTestPV$    && $\claudePythonMHTestPI$     & $\claudePythonMHTestPV$  \\
          & Gemini 1.5 Pro   && $\mathbf{\geminiIproPythonSLTestPI}$ & $\underline{\geminiIproPythonSLTestPV}$ && $\geminiIproPythonSHTestPI$ & $\geminiIproPythonSHTestPV$ && $\underline{\geminiIproPythonMHTestPI}$ & $\underline{\geminiIproPythonMHTestPV}$ \\
          & Gemini 2.0 Flash &&$\underline{\geminiIIflashPythonSLTestPI}$&$\geminiIIflashPythonSLTestPV$&&$\mathbf{\geminiIIflashPythonSHTestPI}$&$\underline{\geminiIIflashPythonSHTestPV}$&&$\underline{\geminiIIflashPythonMHTestPI}$&$\underline{\geminiIIflashPythonMHTestPV}$ \\
          & o3-mini          && $\oIIIminiPythonSLTestPI$   & $\oIIIminiPythonSLTestPV$  && $\underline{\oIIIminiPythonSHTestPI}$   & $\oIIIminiPythonSHTestPV$  &&  $\mathbf{\oIIIminiPythonMHTestPI}$   & $\underline{\oIIIminiPythonMHTestPV}$     \\
          & \deepseekROneDist && $\underline{\RIPythonSLTestPI}$         & $\mathbf{\RIPythonSLTestPV}$        && $\mathbf{\RIPythonSHTestPI}$        &  $\mathbf{\RIPythonSHTestPV}$       && $\underline{\RIPythonMHTestPI}$        & $\mathbf{\RIPythonMHTestPV}$    \\
          & Qwen 2.5 Coder   && $\qwenPythonSLTestPI$      & $\qwenPythonSLTestPV$      && $\qwenPythonSHTestPI$      & $\qwenPythonSHTestPV$      && $\qwenPythonMHTestPI$      & $\qwenPythonMHTestPV$  \\  \cmidrule{1-2} \cmidrule{4-5} \cmidrule{7-8} \cmidrule{10-11}
BugsPHP   & Claude 3.7 Sonnet&& $\underline{\claudePHPSLTestPI}$       & $\underline{\claudePHPSLTestPV}$       && $\mathbf{\claudePHPSHTestPI}$       & $\underline{\claudePHPSHTestPV}$       &&  $\underline{\claudePHPMHTestPI}$       & $\underline{\claudePHPMHTestPV}$  \\
          & Gemini 1.5 Pro   && $\geminiIproPHPSLTestPI$   & $\geminiIproPHPSLTestPV$   && $\geminiIproPHPSHTestPI$   & $\geminiIproPHPSHTestPV$   &&  $\underline{\geminiIproPHPMHTestPI}$   & $\underline{\geminiIproPHPMHTestPV}$    \\
          & Gemini 2.0 Flash && $\underline{\geminiIIflashPHPSLTestPI}$& $\underline{\geminiIIflashPHPSLTestPV}$&& $\geminiIIflashPHPSHTestPI$& $\geminiIIflashPHPSHTestPV$&& $\geminiIIflashPHPMHTestPI$&$\geminiIIflashPHPMHTestPV$ \\
          & o3-mini          && $\mathbf{\oIIIminiPHPSLTestPI}$     & $\mathbf{\oIIIminiPHPSLTestPV}$     && $\underline{\oIIIminiPHPSHTestPI}$     & $\underline{\oIIIminiPHPSHTestPV}$     && $\mathbf{\oIIIminiPHPMHTestPI}$     & $\underline{\oIIIminiPHPMHTestPV}$   \\
          & \deepseekROneDist && $\underline{\RIPHPSLTestPI}$           & $\underline{\RIPHPSLTestPV}$           && $\underline{\RIPHPSHTestPI}$           & $\mathbf{\RIPHPSHTestPV}$           &&  $\mathbf{\RIPHPMHTestPI}$          &  $\mathbf{\RIPHPMHTestPV}$             \\
          & Qwen 2.5 Coder   && $\qwenPHPSLTestPI$         & $\qwenPHPSLTestPV$         && $\qwenPHPSHTestPI$         & $\qwenPHPSHTestPV$         && $\qwenPHPMHTestPI$         & $\qwenPHPMHTestPV$    \\ \cmidrule{1-2} \cmidrule{4-5} \cmidrule{7-8} \cmidrule{10-11}
   Average& &&  $\averageSLTestPI$      & $\averageSLTestPV$       &&  $\averageSHTestPI$      &   $\averageSHTestPV$    &&  $\averageMHTestPI$       & $\averageMHTestPV$\\ \bottomrule
\end{tabular}
\label{tab:bug_complexities_test}
\end{table*}

\subsection{RQ4: open vs closed Models}
Unsurprisingly, the accuracy in most experiments is significantly higher on average for closed models, which are likely trained on larger and higher-quality datasets, are typically larger in parameter count, and may have been subject to more extensive fine-tuning on failure cases compared to open models. 
To get an impression of how open models and closed models progress over time, we plot the quality of program repair against the models' release dates. As a quality measure, we use the pass@5 rate (see Table~\ref{tab:all_models}), averaged over all datasets. 
For OpenAI's models, which undergo continuous development, we use the exact date of the version used in our experiments (for GPT-4o: Nov 11 2024, for o3-mini: Jan 31 2025).

\begin{figure}[t]
    \centering
    \includegraphics[width=0.7\linewidth]{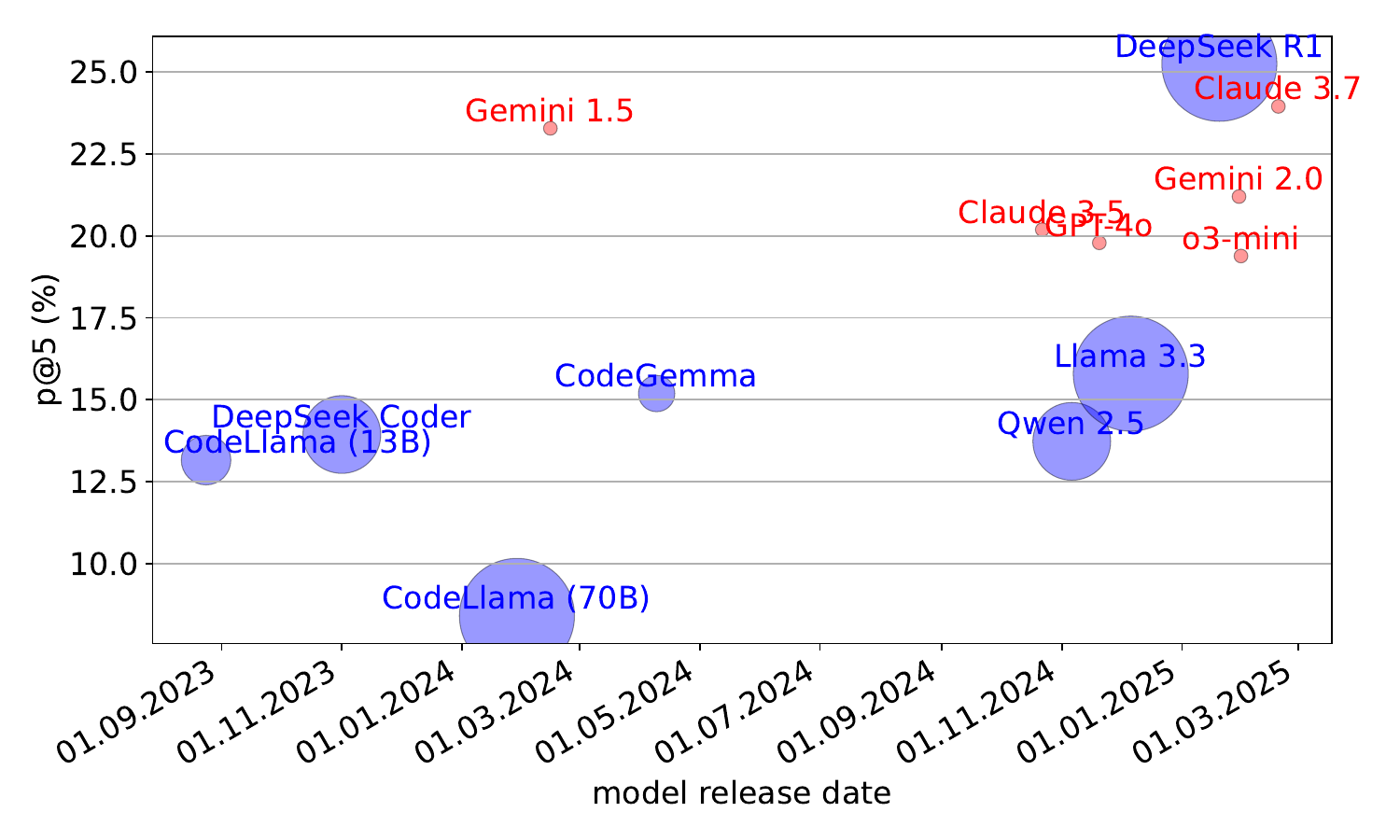}
    \caption{The accuracy of program repair (pass@5, base prompt, macro-averaged over all $4$ benchmarks), plotted against models' release dates. For open models (blue), circle size indicates the respective model's size. Open models are catching up to closed models (red), and have even surpassed them with \deepseekROneDist's release.}
    \label{fig:open-vs-closed}
    \Description{A plot of pass@5 accuracy of open vs closed models in relation to their release date.}
\end{figure}

Figure \ref{fig:open-vs-closed} shows the resulting graph: While a certain stagnation for closed models can be observed, there is a positive trend for open models, with \deepseekROneDist in particular showing strong improvements compared to its predecessors. These findings do not only underline that reasoning models are a particularly interesting future direction for program repair, but also indicate that open models are becoming a viable alternative for software development practitioners, allowing them to avoid vendor lock-in and increasing digital souvereignty.


%% file: values_java.tex
\newcommand{\claudeJavaallSHBASICPI }{ 22.69\% }
\newcommand{\claudeJavaallSHBASICPIII }{ 31.50\% }
\newcommand{\claudeJavaallSHBASICPV }{ 34.90\% }
\newcommand{\claudeHaikuJavaallSHBASICPI }{ 17.74\% }
\newcommand{\claudeHaikuJavaallSHBASICPIII }{ 23.81\% }
\newcommand{\claudeHaikuJavaallSHBASICPV }{ 26.50\% }
\newcommand{\gemmaJavaallSHBASICPI }{ 6.70\% }
\newcommand{\gemmaJavaallSHBASICPIII }{ 12.85\% }
\newcommand{\gemmaJavaallSHBASICPV }{ 16.32\% }
\newcommand{\cLlamaSJavaallSHBASICPI }{ 4.27\% }
\newcommand{\cLlamaSJavaallSHBASICPIII }{ 9.20\% }
\newcommand{\cLlamaSJavaallSHBASICPV }{ 12.50\% }
\newcommand{\LlamaIIIJavaallSHBASICPI }{ 15.12\% }
\newcommand{\LlamaIIIJavaallSHBASICPIII }{ 19.80\% }
\newcommand{\LlamaIIIJavaallSHBASICPV }{ 21.84\% }
\newcommand{\CLlamaLJavaallSHBASICPI }{ 1.52\% }
\newcommand{\CLlamaLJavaallSHBASICPIII }{ 3.55\% }
\newcommand{\CLlamaLJavaallSHBASICPV }{ 4.86\% }
\newcommand{\DSCoderJavaallSHBASICPI }{ 6.34\% }
\newcommand{\DSCoderJavaallSHBASICPIII }{ 12.18\% }
\newcommand{\DSCoderJavaallSHBASICPV }{ 15.61\% }
\newcommand{\gptIVoJavaallSHBASICPI }{ 16.33\% }
\newcommand{\gptIVoJavaallSHBASICPIII }{ 24.56\% }
\newcommand{\gptIVoJavaallSHBASICPV }{ 27.77\% }
\newcommand{\qwenJavaallSHBASICPI }{ 14.67\% }
\newcommand{\qwenJavaallSHBASICPIII }{ 20.26\% }
\newcommand{\qwenJavaallSHBASICPV }{ 22.30\% }
\newcommand{\RIJavaallSHBASICPI }{ 17.57\% }
\newcommand{\RIJavaallSHBASICPIII }{ 26.84\% }
\newcommand{\RIJavaallSHBASICPV }{ 31.35\% }
\newcommand{\geminiIIflashJavaallSHBASICPI }{ 18.83\% }
\newcommand{\geminiIIflashJavaallSHBASICPIII }{ 24.74\% }
\newcommand{\geminiIIflashJavaallSHBASICPV }{ 27.05\% }
\newcommand{\geminiIproJavaallSHBASICPI }{ 20.45\% }
\newcommand{\geminiIproJavaallSHBASICPIII }{ 30.13\% }
\newcommand{\geminiIproJavaallSHBASICPV }{ 33.83\% }
\newcommand{\oIIIminiJavaallSHBASICPI }{ 19.95\% }
\newcommand{\oIIIminiJavaallSHBASICPIII }{ 25.49\% }
\newcommand{\oIIIminiJavaallSHBASICPV }{ 27.96\% }
\newcommand{\claudeJavaBASICPI }{ 19.02\% }
\newcommand{\claudeJavaBASICPIII }{ 28.09\% }
\newcommand{\claudeJavaBASICPV }{ 32.04\% }
\newcommand{\qwenJavaBASICPI }{ 7.41\% }
\newcommand{\qwenJavaBASICPIII }{ 11.64\% }
\newcommand{\qwenJavaBASICPV }{ 12.84\% }
\newcommand{\RIJavaBASICPI }{ 17.16\% }
\newcommand{\RIJavaBASICPIII }{ 25.75\% }
\newcommand{\RIJavaBASICPV }{ 29.82\% }
\newcommand{\geminiIIflashJavaBASICPI }{ 10.27\% }
\newcommand{\geminiIIflashJavaBASICPIII }{ 14.05\% }
\newcommand{\geminiIIflashJavaBASICPV }{ 15.84\% }
\newcommand{\geminiIproJavaBASICPI }{ 19.36\% }
\newcommand{\geminiIproJavaBASICPIII }{ 26.38\% }
\newcommand{\geminiIproJavaBASICPV }{ 28.78\% }
\newcommand{\oIIIminiJavaBASICPI }{ 15.02\% }
\newcommand{\oIIIminiJavaBASICPIII }{ 21.06\% }
\newcommand{\oIIIminiJavaBASICPV }{ 23.61\% }
\newcommand{\claudeJavaSLBASICPI }{ 24.10\% }
\newcommand{\claudeJavaSLBASICPIII }{ 32.44\% }
\newcommand{\claudeJavaSLBASICPV }{ 36.95\% }
\newcommand{\qwenJavaSLBASICPI }{ 13.33\% }
\newcommand{\qwenJavaSLBASICPIII }{ 15.00\% }
\newcommand{\qwenJavaSLBASICPV }{ 16.67\% }
\newcommand{\RIJavaSLBASICPI }{ 21.87\% }
\newcommand{\RIJavaSLBASICPIII }{ 34.86\% }
\newcommand{\RIJavaSLBASICPV }{ 41.39\% }
\newcommand{\geminiIIflashJavaSLBASICPI }{ 12.98\% }
\newcommand{\geminiIIflashJavaSLBASICPIII }{ 17.41\% }
\newcommand{\geminiIIflashJavaSLBASICPV }{ 20.11\% }
\newcommand{\geminiIproJavaSLBASICPI }{ 26.67\% }
\newcommand{\geminiIproJavaSLBASICPIII }{ 38.00\% }
\newcommand{\geminiIproJavaSLBASICPV }{ 41.27\% }
\newcommand{\oIIIminiJavaSLBASICPI }{ 18.77\% }
\newcommand{\oIIIminiJavaSLBASICPIII }{ 24.38\% }
\newcommand{\oIIIminiJavaSLBASICPV }{ 27.92\% }
\newcommand{\claudeJavaSHBASICPI }{ 19.13\% }
\newcommand{\claudeJavaSHBASICPIII }{ 29.48\% }
\newcommand{\claudeJavaSHBASICPV }{ 32.18\% }
\newcommand{\qwenJavaSHBASICPI }{ 10.26\% }
\newcommand{\qwenJavaSHBASICPIII }{ 15.05\% }
\newcommand{\qwenJavaSHBASICPV }{ 15.38\% }
\newcommand{\RIJavaSHBASICPI }{ 12.38\% }
\newcommand{\RIJavaSHBASICPIII }{ 16.14\% }
\newcommand{\RIJavaSHBASICPV }{ 17.46\% }
\newcommand{\geminiIIflashJavaSHBASICPI }{ 5.56\% }
\newcommand{\geminiIIflashJavaSHBASICPIII }{ 5.56\% }
\newcommand{\geminiIIflashJavaSHBASICPV }{ 5.56\% }
\newcommand{\geminiIproJavaSHBASICPI }{ 12.17\% }
\newcommand{\geminiIproJavaSHBASICPIII }{ 21.11\% }
\newcommand{\geminiIproJavaSHBASICPV }{ 24.62\% }
\newcommand{\oIIIminiJavaSHBASICPI }{ 15.07\% }
\newcommand{\oIIIminiJavaSHBASICPIII }{ 21.46\% }
\newcommand{\oIIIminiJavaSHBASICPV }{ 24.16\% }
\newcommand{\claudeJavaMHBASICPI }{ 15.96\% }
\newcommand{\claudeJavaMHBASICPIII }{ 25.01\% }
\newcommand{\claudeJavaMHBASICPV }{ 29.44\% }
\newcommand{\qwenJavaMHBASICPI }{ 5.53\% }
\newcommand{\qwenJavaMHBASICPIII }{ 10.20\% }
\newcommand{\qwenJavaMHBASICPV }{ 11.61\% }
\newcommand{\geminiIIflashJavaMHBASICPI }{ 11.48\% }
\newcommand{\geminiIIflashJavaMHBASICPIII }{ 16.90\% }
\newcommand{\geminiIIflashJavaMHBASICPV }{ 19.16\% }
\newcommand{\geminiIproJavaMHBASICPI }{ 18.71\% }
\newcommand{\geminiIproJavaMHBASICPIII }{ 20.64\% }
\newcommand{\geminiIproJavaMHBASICPV }{ 21.51\% }
\newcommand{\oIIIminiJavaMHBASICPI }{ 12.59\% }
\newcommand{\oIIIminiJavaMHBASICPIII }{ 18.89\% }
\newcommand{\oIIIminiJavaMHBASICPV }{ 20.68\% }
\newcommand{\claudeJavaTestPI }{ 45.33\% }
\newcommand{\claudeJavaTestPIII }{ 55.68\% }
\newcommand{\claudeJavaTestPV }{ 59.16\% }
\newcommand{\qwenJavaTestPI }{ 16.13\% }
\newcommand{\qwenJavaTestPIII }{ 25.64\% }
\newcommand{\qwenJavaTestPV }{ 28.86\% }
\newcommand{\RIJavaTestPI }{ 32.81\% }
\newcommand{\RIJavaTestPIII }{ 43.22\% }
\newcommand{\RIJavaTestPV }{ 47.70\% }
\newcommand{\geminiIIflashJavaTestPI }{ 33.56\% }
\newcommand{\geminiIIflashJavaTestPIII }{ 39.70\% }
\newcommand{\geminiIIflashJavaTestPV }{ 41.59\% }
\newcommand{\geminiIproJavaTestPI }{ 33.86\% }
\newcommand{\geminiIproJavaTestPIII }{ 40.14\% }
\newcommand{\geminiIproJavaTestPV }{ 41.75\% }
\newcommand{\oIIIminiJavaTestPI }{ 48.50\% }
\newcommand{\oIIIminiJavaTestPIII }{ 59.31\% }
\newcommand{\oIIIminiJavaTestPV }{ 62.60\% }
\newcommand{\claudeJavaSLTestPI }{ 56.54\% }
\newcommand{\claudeJavaSLTestPIII }{ 65.87\% }
\newcommand{\claudeJavaSLTestPV }{ 68.49\% }
\newcommand{\qwenJavaSLTestPI }{ 19.75\% }
\newcommand{\qwenJavaSLTestPIII }{ 35.90\% }
\newcommand{\qwenJavaSLTestPV }{ 40.84\% }
\newcommand{\RIJavaSLTestPI }{ 50.26\% }
\newcommand{\RIJavaSLTestPIII }{ 62.56\% }
\newcommand{\RIJavaSLTestPV }{ 67.00\% }
\newcommand{\geminiIIflashJavaSLTestPI }{ 44.62\% }
\newcommand{\geminiIIflashJavaSLTestPIII }{ 46.88\% }
\newcommand{\geminiIIflashJavaSLTestPV }{ 47.44\% }
\newcommand{\geminiIproJavaSLTestPI }{ 42.93\% }
\newcommand{\geminiIproJavaSLTestPIII }{ 51.25\% }
\newcommand{\geminiIproJavaSLTestPV }{ 53.88\% }
\newcommand{\oIIIminiJavaSLTestPI }{ 58.93\% }
\newcommand{\oIIIminiJavaSLTestPIII }{ 66.43\% }
\newcommand{\oIIIminiJavaSLTestPV }{ 69.37\% }
\newcommand{\claudeJavaSHTestPI }{ 44.24\% }
\newcommand{\claudeJavaSHTestPIII }{ 50.16\% }
\newcommand{\claudeJavaSHTestPV }{ 52.90\% }
\newcommand{\qwenJavaSHTestPI }{ 13.64\% }
\newcommand{\qwenJavaSHTestPIII }{ 23.50\% }
\newcommand{\qwenJavaSHTestPV }{ 26.96\% }
\newcommand{\RIJavaSHTestPI }{ 25.15\% }
\newcommand{\RIJavaSHTestPIII }{ 33.23\% }
\newcommand{\RIJavaSHTestPV }{ 37.30\% }
\newcommand{\geminiIIflashJavaSHTestPI }{ 25.71\% }
\newcommand{\geminiIIflashJavaSHTestPIII }{ 30.65\% }
\newcommand{\geminiIIflashJavaSHTestPV }{ 32.04\% }
\newcommand{\geminiIproJavaSHTestPI }{ 28.67\% }
\newcommand{\geminiIproJavaSHTestPIII }{ 34.38\% }
\newcommand{\geminiIproJavaSHTestPV }{ 34.96\% }
\newcommand{\oIIIminiJavaSHTestPI }{ 48.67\% }
\newcommand{\oIIIminiJavaSHTestPIII }{ 61.40\% }
\newcommand{\oIIIminiJavaSHTestPV }{ 66.15\% }
\newcommand{\claudeJavaMHTestPI }{ 39.28\% }
\newcommand{\claudeJavaMHTestPIII }{ 52.34\% }
\newcommand{\claudeJavaMHTestPV }{ 56.68\% }
\newcommand{\qwenJavaMHTestPI }{ 15.50\% }
\newcommand{\qwenJavaMHTestPIII }{ 20.88\% }
\newcommand{\qwenJavaMHTestPV }{ 22.99\% }
\newcommand{\RIJavaMHTestPI }{ 26.03\% }
\newcommand{\RIJavaMHTestPIII }{ 36.48\% }
\newcommand{\RIJavaMHTestPV }{ 41.21\% }
\newcommand{\geminiIIflashJavaMHTestPI }{ 31.38\% }
\newcommand{\geminiIIflashJavaMHTestPIII }{ 40.76\% }
\newcommand{\geminiIIflashJavaMHTestPV }{ 43.78\% }
\newcommand{\geminiIproJavaMHTestPI }{ 31.75\% }
\newcommand{\geminiIproJavaMHTestPIII }{ 37.22\% }
\newcommand{\geminiIproJavaMHTestPV }{ 38.76\% }
\newcommand{\oIIIminiJavaMHTestPI }{ 42.75\% }
\newcommand{\oIIIminiJavaMHTestPIII }{ 54.53\% }
\newcommand{\oIIIminiJavaMHTestPV }{ 57.37\% }
\newcommand{\claudeJavaPerfectLLPI }{ 27.78\% }
\newcommand{\claudeJavaPerfectLLPIII }{ 35.41\% }
\newcommand{\claudeJavaPerfectLLPV }{ 38.13\% }
\newcommand{\qwenJavaPerfectLLPI }{ 11.81\% }
\newcommand{\qwenJavaPerfectLLPIII }{ 16.04\% }
\newcommand{\qwenJavaPerfectLLPV }{ 17.96\% }
\newcommand{\RIJavaPerfectLLPI }{ 18.00\% }
\newcommand{\RIJavaPerfectLLPIII }{ 26.21\% }
\newcommand{\RIJavaPerfectLLPV }{ 29.67\% }
\newcommand{\geminiIIflashJavaPerfectLLPI }{ 14.25\% }
\newcommand{\geminiIIflashJavaPerfectLLPIII }{ 18.68\% }
\newcommand{\geminiIIflashJavaPerfectLLPV }{ 20.87\% }
\newcommand{\geminiIproJavaPerfectLLPI }{ 20.36\% }
\newcommand{\geminiIproJavaPerfectLLPIII }{ 24.85\% }
\newcommand{\geminiIproJavaPerfectLLPV }{ 26.24\% }
\newcommand{\oIIIminiJavaPerfectLLPI }{ 22.81\% }
\newcommand{\oIIIminiJavaPerfectLLPIII }{ 30.54\% }
\newcommand{\oIIIminiJavaPerfectLLPV }{ 33.21\% }
\newcommand{\claudeJavaSLPerfectLLPI }{ 35.31\% }
\newcommand{\claudeJavaSLPerfectLLPIII }{ 44.20\% }
\newcommand{\claudeJavaSLPerfectLLPV }{ 47.46\% }
\newcommand{\qwenJavaSLPerfectLLPI }{ 15.56\% }
\newcommand{\qwenJavaSLPerfectLLPIII }{ 21.38\% }
\newcommand{\qwenJavaSLPerfectLLPV }{ 24.24\% }
\newcommand{\RIJavaSLPerfectLLPI }{ 24.20\% }
\newcommand{\RIJavaSLPerfectLLPIII }{ 35.53\% }
\newcommand{\RIJavaSLPerfectLLPV }{ 39.72\% }
\newcommand{\geminiIIflashJavaSLPerfectLLPI }{ 16.54\% }
\newcommand{\geminiIIflashJavaSLPerfectLLPIII }{ 22.55\% }
\newcommand{\geminiIIflashJavaSLPerfectLLPV }{ 26.02\% }
\newcommand{\geminiIproJavaSLPerfectLLPI }{ 32.00\% }
\newcommand{\geminiIproJavaSLPerfectLLPIII }{ 39.12\% }
\newcommand{\geminiIproJavaSLPerfectLLPV }{ 39.96\% }
\newcommand{\oIIIminiJavaSLPerfectLLPI }{ 29.74\% }
\newcommand{\oIIIminiJavaSLPerfectLLPIII }{ 36.38\% }
\newcommand{\oIIIminiJavaSLPerfectLLPV }{ 39.04\% }
\newcommand{\claudeJavaSHPerfectLLPI }{ 25.76\% }
\newcommand{\claudeJavaSHPerfectLLPIII }{ 32.07\% }
\newcommand{\claudeJavaSHPerfectLLPV }{ 34.03\% }
\newcommand{\qwenJavaSHPerfectLLPI }{ 10.61\% }
\newcommand{\qwenJavaSHPerfectLLPIII }{ 13.40\% }
\newcommand{\qwenJavaSHPerfectLLPV }{ 13.63\% }
\newcommand{\RIJavaSHPerfectLLPI }{ 13.33\% }
\newcommand{\RIJavaSHPerfectLLPIII }{ 17.10\% }
\newcommand{\RIJavaSHPerfectLLPV }{ 19.26\% }
\newcommand{\geminiIIflashJavaSHPerfectLLPI }{ 7.88\% }
\newcommand{\geminiIIflashJavaSHPerfectLLPIII }{ 9.90\% }
\newcommand{\geminiIIflashJavaSHPerfectLLPV }{ 10.60\% }
\newcommand{\geminiIproJavaSHPerfectLLPI }{ 22.22\% }
\newcommand{\geminiIproJavaSHPerfectLLPIII }{ 32.60\% }
\newcommand{\geminiIproJavaSHPerfectLLPV }{ 33.32\% }
\newcommand{\oIIIminiJavaSHPerfectLLPI }{ 22.86\% }
\newcommand{\oIIIminiJavaSHPerfectLLPIII }{ 36.04\% }
\newcommand{\oIIIminiJavaSHPerfectLLPV }{ 40.25\% }
\newcommand{\claudeJavaMHPerfectLLPI }{ 23.90\% }
\newcommand{\claudeJavaMHPerfectLLPIII }{ 31.41\% }
\newcommand{\claudeJavaMHPerfectLLPV }{ 34.19\% }
\newcommand{\qwenJavaMHPerfectLLPI }{ 10.00\% }
\newcommand{\qwenJavaMHPerfectLLPIII }{ 13.50\% }
\newcommand{\qwenJavaMHPerfectLLPV }{ 15.26\% }
\newcommand{\RIJavaMHPerfectLLPI }{ 16.42\% }
\newcommand{\RIJavaMHPerfectLLPIII }{ 24.97\% }
\newcommand{\RIJavaMHPerfectLLPV }{ 28.64\% }
\newcommand{\geminiIIflashJavaMHPerfectLLPI }{ 16.30\% }
\newcommand{\geminiIIflashJavaMHPerfectLLPIII }{ 21.07\% }
\newcommand{\geminiIIflashJavaMHPerfectLLPV }{ 23.26\% }
\newcommand{\geminiIproJavaMHPerfectLLPI }{ 18.79\% }
\newcommand{\geminiIproJavaMHPerfectLLPIII }{ 22.17\% }
\newcommand{\geminiIproJavaMHPerfectLLPV }{ 23.71\% }
\newcommand{\oIIIminiJavaMHPerfectLLPI }{ 18.60\% }
\newcommand{\oIIIminiJavaMHPerfectLLPIII }{ 24.33\% }
\newcommand{\oIIIminiJavaMHPerfectLLPV }{ 26.24\% }
\newcommand{\claudeJavaFlacocoPI }{ 2.76\% }
\newcommand{\claudeJavaFlacocoPIII }{ 5.02\% }
\newcommand{\claudeJavaFlacocoPV }{ 6.39\% }
\newcommand{\qwenJavaFlacocoPI }{ 1.61\% }
\newcommand{\qwenJavaFlacocoPIII }{ 2.31\% }
\newcommand{\qwenJavaFlacocoPV }{ 2.78\% }
\newcommand{\RIJavaFlacocoPI }{ 0.31\% }
\newcommand{\RIJavaFlacocoPIII }{ 0.84\% }
\newcommand{\RIJavaFlacocoPV }{ 1.26\% }
\newcommand{\geminiIIflashJavaFlacocoPI }{ 2.40\% }
\newcommand{\geminiIIflashJavaFlacocoPIII }{ 3.62\% }
\newcommand{\geminiIIflashJavaFlacocoPV }{ 4.10\% }
\newcommand{\geminiIproJavaFlacocoPI }{ 3.33\% }
\newcommand{\geminiIproJavaFlacocoPIII }{ 5.40\% }
\newcommand{\geminiIproJavaFlacocoPV }{ 6.27\% }
\newcommand{\oIIIminiJavaFlacocoPI }{ 3.64\% }
\newcommand{\oIIIminiJavaFlacocoPIII }{ 6.65\% }
\newcommand{\oIIIminiJavaFlacocoPV }{ 8.16\% }
\newcommand{\claudeJavaSLFlacocoPI }{ 2.96\% }
\newcommand{\claudeJavaSLFlacocoPIII }{ 7.39\% }
\newcommand{\claudeJavaSLFlacocoPV }{ 10.35\% }
\newcommand{\qwenJavaSLFlacocoPI }{ 0.51\% }
\newcommand{\qwenJavaSLFlacocoPIII }{ 1.43\% }
\newcommand{\qwenJavaSLFlacocoPV }{ 2.20\% }
\newcommand{\RIJavaSLFlacocoPI }{ 0.26\% }
\newcommand{\RIJavaSLFlacocoPIII }{ 0.77\% }
\newcommand{\RIJavaSLFlacocoPV }{ 1.28\% }
\newcommand{\geminiIIflashJavaSLFlacocoPI }{ 2.56\% }
\newcommand{\geminiIIflashJavaSLFlacocoPIII }{ 4.98\% }
\newcommand{\geminiIIflashJavaSLFlacocoPV }{ 6.02\% }
\newcommand{\geminiIproJavaSLFlacocoPI }{ 3.08\% }
\newcommand{\geminiIproJavaSLFlacocoPIII }{ 7.11\% }
\newcommand{\geminiIproJavaSLFlacocoPV }{ 9.35\% }
\newcommand{\oIIIminiJavaSLFlacocoPI }{ 6.67\% }
\newcommand{\oIIIminiJavaSLFlacocoPIII }{ 14.55\% }
\newcommand{\oIIIminiJavaSLFlacocoPV }{ 18.35\% }
\newcommand{\claudeJavaSHFlacocoPI }{ 4.76\% }
\newcommand{\claudeJavaSHFlacocoPIII }{ 4.76\% }
\newcommand{\claudeJavaSHFlacocoPV }{ 4.76\% }
\newcommand{\qwenJavaSHFlacocoPI }{ 4.55\% }
\newcommand{\qwenJavaSHFlacocoPIII }{ 4.55\% }
\newcommand{\qwenJavaSHFlacocoPV }{ 4.55\% }
\newcommand{\RIJavaSHFlacocoPI }{ 0.00\% }
\newcommand{\RIJavaSHFlacocoPIII }{ 0.00\% }
\newcommand{\RIJavaSHFlacocoPV }{ 0.00\% }
\newcommand{\geminiIIflashJavaSHFlacocoPI }{ 4.76\% }
\newcommand{\geminiIIflashJavaSHFlacocoPIII }{ 4.76\% }
\newcommand{\geminiIIflashJavaSHFlacocoPV }{ 4.76\% }
\newcommand{\geminiIproJavaSHFlacocoPI }{ 4.76\% }
\newcommand{\geminiIproJavaSHFlacocoPIII }{ 4.76\% }
\newcommand{\geminiIproJavaSHFlacocoPV }{ 4.76\% }
\newcommand{\oIIIminiJavaSHFlacocoPI }{ 4.76\% }
\newcommand{\oIIIminiJavaSHFlacocoPIII }{ 4.76\% }
\newcommand{\oIIIminiJavaSHFlacocoPV }{ 4.76\% }
\newcommand{\claudeJavaMHFlacocoPI }{ 1.54\% }
\newcommand{\claudeJavaMHFlacocoPIII }{ 3.52\% }
\newcommand{\claudeJavaMHFlacocoPV }{ 4.52\% }
\newcommand{\qwenJavaMHFlacocoPI }{ 0.68\% }
\newcommand{\qwenJavaMHFlacocoPIII }{ 1.63\% }
\newcommand{\qwenJavaMHFlacocoPV }{ 2.17\% }
\newcommand{\RIJavaMHFlacocoPI }{ 0.53\% }
\newcommand{\RIJavaMHFlacocoPIII }{ 1.36\% }
\newcommand{\RIJavaMHFlacocoPV }{ 1.94\% }
\newcommand{\geminiIIflashJavaMHFlacocoPI }{ 1.03\% }
\newcommand{\geminiIIflashJavaMHFlacocoPIII }{ 2.09\% }
\newcommand{\geminiIIflashJavaMHFlacocoPV }{ 2.46\% }
\newcommand{\geminiIproJavaMHFlacocoPI }{ 2.74\% }
\newcommand{\geminiIproJavaMHFlacocoPIII }{ 4.60\% }
\newcommand{\geminiIproJavaMHFlacocoPV }{ 5.02\% }
\newcommand{\oIIIminiJavaMHFlacocoPI }{ 1.03\% }
\newcommand{\oIIIminiJavaMHFlacocoPIII }{ 2.40\% }
\newcommand{\oIIIminiJavaMHFlacocoPV }{ 3.20\% }

%% file: values_js.tex
\newcommand{\claudeJSallSHBASICPI }{ 11.71\% }
\newcommand{\claudeJSallSHBASICPIII }{ 19.83\% }
\newcommand{\claudeJSallSHBASICPV }{ 24.23\% }
\newcommand{\claudeHaikuJSallSHBASICPI }{ 14.48\% }
\newcommand{\claudeHaikuJSallSHBASICPIII }{ 21.89\% }
\newcommand{\claudeHaikuJSallSHBASICPV }{ 24.14\% }
\newcommand{\gemmaJSallSHBASICPI }{ 6.50\% }
\newcommand{\gemmaJSallSHBASICPIII }{ 14.41\% }
\newcommand{\gemmaJSallSHBASICPV }{ 19.88\% }
\newcommand{\cLlamaSJSallSHBASICPI }{ 6.43\% }
\newcommand{\cLlamaSJSallSHBASICPIII }{ 15.55\% }
\newcommand{\cLlamaSJSallSHBASICPV }{ 21.43\% }
\newcommand{\LlamaIIIJSallSHBASICPI }{ 9.09\% }
\newcommand{\LlamaIIIJSallSHBASICPIII }{ 12.96\% }
\newcommand{\LlamaIIIJSallSHBASICPV }{ 14.96\% }
\newcommand{\DSCoderJSallSHBASICPI }{ 4.28\% }
\newcommand{\DSCoderJSallSHBASICPIII }{ 9.29\% }
\newcommand{\DSCoderJSallSHBASICPV }{ 12.03\% }
\newcommand{\gptIVoJSallSHBASICPI }{ 10.57\% }
\newcommand{\gptIVoJSallSHBASICPIII }{ 19.33\% }
\newcommand{\gptIVoJSallSHBASICPV }{ 23.53\% }
\newcommand{\qwenJSallSHBASICPI }{ 7.93\% }
\newcommand{\qwenJSallSHBASICPIII }{ 13.10\% }
\newcommand{\qwenJSallSHBASICPV }{ 15.71\% }
\newcommand{\RIJSallSHBASICPI }{ 9.55\% }
\newcommand{\RIJSallSHBASICPIII }{ 17.21\% }
\newcommand{\RIJSallSHBASICPV }{ 21.66\% }
\newcommand{\geminiIIflashJSallSHBASICPI }{ 10.34\% }
\newcommand{\geminiIIflashJSallSHBASICPIII }{ 14.59\% }
\newcommand{\geminiIIflashJSallSHBASICPV }{ 16.15\% }
\newcommand{\geminiIproJSallSHBASICPI }{ 11.14\% }
\newcommand{\geminiIproJSallSHBASICPIII }{ 17.11\% }
\newcommand{\geminiIproJSallSHBASICPV }{ 19.73\% }
\newcommand{\oIIIminiJSallSHBASICPI }{ 9.20\% }
\newcommand{\oIIIminiJSallSHBASICPIII }{ 14.90\% }
\newcommand{\oIIIminiJSallSHBASICPV }{ 18.10\% }
\newcommand{\claudeJSBASICPI }{ 8.39\% }
\newcommand{\claudeJSBASICPIII }{ 15.22\% }
\newcommand{\claudeJSBASICPV }{ 19.12\% }
\newcommand{\qwenJSBASICPI }{ 5.11\% }
\newcommand{\qwenJSBASICPIII }{ 8.73\% }
\newcommand{\qwenJSBASICPV }{ 10.73\% }
\newcommand{\RIJSBASICPI }{ 7.71\% }
\newcommand{\RIJSBASICPIII }{ 14.68\% }
\newcommand{\RIJSBASICPV }{ 18.95\% }
\newcommand{\geminiIIflashJSBASICPI }{ 9.10\% }
\newcommand{\geminiIIflashJSBASICPIII }{ 12.80\% }
\newcommand{\geminiIIflashJSBASICPV }{ 13.90\% }
\newcommand{\geminiIproJSBASICPI }{ 8.39\% }
\newcommand{\geminiIproJSBASICPIII }{ 12.79\% }
\newcommand{\geminiIproJSBASICPV }{ 14.70\% }
\newcommand{\oIIIminiJSBASICPI }{ 6.32\% }
\newcommand{\oIIIminiJSBASICPIII }{ 10.59\% }
\newcommand{\oIIIminiJSBASICPV }{ 13.01\% }
\newcommand{\claudeJSSLBASICPI }{ 18.49\% }
\newcommand{\claudeJSSLBASICPIII }{ 30.82\% }
\newcommand{\claudeJSSLBASICPV }{ 37.97\% }
\newcommand{\qwenJSSLBASICPI }{ 13.58\% }
\newcommand{\qwenJSSLBASICPIII }{ 20.69\% }
\newcommand{\qwenJSSLBASICPV }{ 23.99\% }
\newcommand{\RIJSSLBASICPI }{ 16.15\% }
\newcommand{\RIJSSLBASICPIII }{ 27.96\% }
\newcommand{\RIJSSLBASICPV }{ 34.88\% }
\newcommand{\geminiIIflashJSSLBASICPI }{ 14.57\% }
\newcommand{\geminiIIflashJSSLBASICPIII }{ 20.77\% }
\newcommand{\geminiIIflashJSSLBASICPV }{ 23.65\% }
\newcommand{\geminiIproJSSLBASICPI }{ 17.78\% }
\newcommand{\geminiIproJSSLBASICPIII }{ 23.88\% }
\newcommand{\geminiIproJSSLBASICPV }{ 26.04\% }
\newcommand{\oIIIminiJSSLBASICPI }{ 15.06\% }
\newcommand{\oIIIminiJSSLBASICPIII }{ 23.12\% }
\newcommand{\oIIIminiJSSLBASICPV }{ 27.46\% }
\newcommand{\claudeJSSHBASICPI }{ 5.81\% }
\newcommand{\claudeJSSHBASICPIII }{ 10.27\% }
\newcommand{\claudeJSSHBASICPV }{ 12.27\% }
\newcommand{\qwenJSSHBASICPI }{ 3.01\% }
\newcommand{\qwenJSSHBASICPIII }{ 6.49\% }
\newcommand{\qwenJSSHBASICPV }{ 8.50\% }
\newcommand{\RIJSSHBASICPI }{ 4.00\% }
\newcommand{\RIJSSHBASICPIII }{ 8.18\% }
\newcommand{\RIJSSHBASICPV }{ 10.56\% }
\newcommand{\geminiIIflashJSSHBASICPI }{ 6.67\% }
\newcommand{\geminiIIflashJSSHBASICPIII }{ 9.21\% }
\newcommand{\geminiIIflashJSSHBASICPV }{ 9.62\% }
\newcommand{\geminiIproJSSHBASICPI }{ 5.35\% }
\newcommand{\geminiIproJSSHBASICPIII }{ 11.21\% }
\newcommand{\geminiIproJSSHBASICPV }{ 14.24\% }
\newcommand{\oIIIminiJSSHBASICPI }{ 4.09\% }
\newcommand{\oIIIminiJSSHBASICPIII }{ 7.74\% }
\newcommand{\oIIIminiJSSHBASICPV }{ 9.95\% }
\newcommand{\claudeJSMHBASICPI }{ 3.33\% }
\newcommand{\claudeJSMHBASICPIII }{ 8.17\% }
\newcommand{\claudeJSMHBASICPV }{ 11.33\% }
\newcommand{\qwenJSMHBASICPI }{ 0.68\% }
\newcommand{\qwenJSMHBASICPIII }{ 1.89\% }
\newcommand{\qwenJSMHBASICPV }{ 2.93\% }
\newcommand{\RIJSMHBASICPI }{ 5.37\% }
\newcommand{\RIJSMHBASICPIII }{ 11.46\% }
\newcommand{\RIJSMHBASICPV }{ 15.48\% }
\newcommand{\geminiIIflashJSMHBASICPI }{ 7.19\% }
\newcommand{\geminiIIflashJSMHBASICPIII }{ 10.06\% }
\newcommand{\geminiIIflashJSMHBASICPV }{ 10.47\% }
\newcommand{\geminiIproJSMHBASICPI }{ 4.21\% }
\newcommand{\geminiIproJSMHBASICPIII }{ 6.20\% }
\newcommand{\geminiIproJSMHBASICPV }{ 7.01\% }
\newcommand{\oIIIminiJSMHBASICPI }{ 1.93\% }
\newcommand{\oIIIminiJSMHBASICPIII }{ 4.01\% }
\newcommand{\oIIIminiJSMHBASICPV }{ 5.24\% }
\newcommand{\claudeJSPerfectLLPI }{ 20.90\% }
\newcommand{\claudeJSPerfectLLPIII }{ 26.42\% }
\newcommand{\claudeJSPerfectLLPV }{ 29.03\% }
\newcommand{\qwenJSPerfectLLPI }{ 8.96\% }
\newcommand{\qwenJSPerfectLLPIII }{ 13.68\% }
\newcommand{\qwenJSPerfectLLPV }{ 16.12\% }
\newcommand{\RIJSPerfectLLPI }{ 12.58\% }
\newcommand{\RIJSPerfectLLPIII }{ 23.77\% }
\newcommand{\RIJSPerfectLLPV }{ 29.88\% }
\newcommand{\geminiIIflashJSPerfectLLPI }{ 18.53\% }
\newcommand{\geminiIIflashJSPerfectLLPIII }{ 22.07\% }
\newcommand{\geminiIIflashJSPerfectLLPV }{ 23.31\% }
\newcommand{\geminiIproJSPerfectLLPI }{ 20.35\% }
\newcommand{\geminiIproJSPerfectLLPIII }{ 25.16\% }
\newcommand{\geminiIproJSPerfectLLPV }{ 27.48\% }
\newcommand{\oIIIminiJSPerfectLLPI }{ 18.92\% }
\newcommand{\oIIIminiJSPerfectLLPIII }{ 25.56\% }
\newcommand{\oIIIminiJSPerfectLLPV }{ 28.89\% }
\newcommand{\claudeJSSLPerfectLLPI }{ 37.04\% }
\newcommand{\claudeJSSLPerfectLLPIII }{ 42.63\% }
\newcommand{\claudeJSSLPerfectLLPV }{ 45.14\% }
\newcommand{\qwenJSSLPerfectLLPI }{ 12.35\% }
\newcommand{\qwenJSSLPerfectLLPIII }{ 21.17\% }
\newcommand{\qwenJSSLPerfectLLPV }{ 26.68\% }
\newcommand{\RIJSSLPerfectLLPI }{ 16.81\% }
\newcommand{\RIJSSLPerfectLLPIII }{ 34.62\% }
\newcommand{\RIJSSLPerfectLLPV }{ 45.08\% }
\newcommand{\geminiIIflashJSSLPerfectLLPI }{ 31.11\% }
\newcommand{\geminiIIflashJSSLPerfectLLPIII }{ 35.08\% }
\newcommand{\geminiIIflashJSSLPerfectLLPV }{ 36.05\% }
\newcommand{\geminiIproJSSLPerfectLLPI }{ 34.07\% }
\newcommand{\geminiIproJSSLPerfectLLPIII }{ 39.79\% }
\newcommand{\geminiIproJSSLPerfectLLPV }{ 42.90\% }
\newcommand{\oIIIminiJSSLPerfectLLPI }{ 35.90\% }
\newcommand{\oIIIminiJSSLPerfectLLPIII }{ 43.96\% }
\newcommand{\oIIIminiJSSLPerfectLLPV }{ 48.30\% }
\newcommand{\claudeJSSHPerfectLLPI }{ 15.27\% }
\newcommand{\claudeJSSHPerfectLLPIII }{ 19.58\% }
\newcommand{\claudeJSSHPerfectLLPV }{ 21.72\% }
\newcommand{\qwenJSSHPerfectLLPI }{ 10.11\% }
\newcommand{\qwenJSSHPerfectLLPIII }{ 16.27\% }
\newcommand{\qwenJSSHPerfectLLPV }{ 18.39\% }
\newcommand{\RIJSSHPerfectLLPI }{ 10.22\% }
\newcommand{\RIJSSHPerfectLLPIII }{ 19.09\% }
\newcommand{\RIJSSHPerfectLLPV }{ 23.99\% }
\newcommand{\geminiIIflashJSSHPerfectLLPI }{ 10.97\% }
\newcommand{\geminiIIflashJSSHPerfectLLPIII }{ 12.93\% }
\newcommand{\geminiIIflashJSSHPerfectLLPV }{ 14.25\% }
\newcommand{\geminiIproJSSHPerfectLLPI }{ 20.65\% }
\newcommand{\geminiIproJSSHPerfectLLPIII }{ 25.38\% }
\newcommand{\geminiIproJSSHPerfectLLPV }{ 27.91\% }
\newcommand{\oIIIminiJSSHPerfectLLPI }{ 13.11\% }
\newcommand{\oIIIminiJSSHPerfectLLPIII }{ 17.88\% }
\newcommand{\oIIIminiJSSHPerfectLLPV }{ 19.60\% }
\newcommand{\claudeJSMHPerfectLLPI }{ 14.04\% }
\newcommand{\claudeJSMHPerfectLLPIII }{ 20.49\% }
\newcommand{\claudeJSMHPerfectLLPV }{ 23.55\% }
\newcommand{\qwenJSMHPerfectLLPI }{ 5.61\% }
\newcommand{\qwenJSMHPerfectLLPIII }{ 6.24\% }
\newcommand{\qwenJSMHPerfectLLPV }{ 6.77\% }
\newcommand{\RIJSMHPerfectLLPI }{ 11.85\% }
\newcommand{\RIJSMHPerfectLLPIII }{ 20.73\% }
\newcommand{\RIJSMHPerfectLLPV }{ 25.08\% }
\newcommand{\geminiIIflashJSMHPerfectLLPI }{ 15.68\% }
\newcommand{\geminiIIflashJSMHPerfectLLPIII }{ 20.24\% }
\newcommand{\geminiIIflashJSMHPerfectLLPV }{ 21.61\% }
\newcommand{\geminiIproJSMHPerfectLLPI }{ 10.35\% }
\newcommand{\geminiIproJSMHPerfectLLPIII }{ 14.60\% }
\newcommand{\geminiIproJSMHPerfectLLPV }{ 16.18\% }
\newcommand{\oIIIminiJSMHPerfectLLPI }{ 11.71\% }
\newcommand{\oIIIminiJSMHPerfectLLPIII }{ 18.85\% }
\newcommand{\oIIIminiJSMHPerfectLLPV }{ 22.80\% }
\newcommand{\claudeJSTestPI }{ 42.39\% }
\newcommand{\claudeJSTestPIII }{ 51.94\% }
\newcommand{\claudeJSTestPV }{ 56.34\% }
\newcommand{\qwenJSTestPI }{ 24.00\% }
\newcommand{\qwenJSTestPIII }{ 31.08\% }
\newcommand{\qwenJSTestPV }{ 34.86\% }
\newcommand{\RIJSTestPI }{ 34.79\% }
\newcommand{\RIJSTestPIII }{ 50.32\% }
\newcommand{\RIJSTestPV }{ 56.71\% }
\newcommand{\geminiIIflashJSTestPI }{ 28.40\% }
\newcommand{\geminiIIflashJSTestPIII }{ 36.10\% }
\newcommand{\geminiIIflashJSTestPV }{ 39.05\% }
\newcommand{\geminiIproJSTestPI }{ 26.39\% }
\newcommand{\geminiIproJSTestPIII }{ 30.73\% }
\newcommand{\geminiIproJSTestPV }{ 32.41\% }
\newcommand{\oIIIminiJSTestPI }{ 51.44\% }
\newcommand{\oIIIminiJSTestPIII }{ 63.79\% }
\newcommand{\oIIIminiJSTestPV }{ 68.00\% }
\newcommand{\claudeJSSLTestPI }{ 52.59\% }
\newcommand{\claudeJSSLTestPIII }{ 61.25\% }
\newcommand{\claudeJSSLTestPV }{ 64.04\% }
\newcommand{\qwenJSSLTestPI }{ 44.69\% }
\newcommand{\qwenJSSLTestPIII }{ 53.76\% }
\newcommand{\qwenJSSLTestPV }{ 58.38\% }
\newcommand{\RIJSSLTestPI }{ 54.78\% }
\newcommand{\RIJSSLTestPIII }{ 72.25\% }
\newcommand{\RIJSSLTestPV }{ 77.68\% }
\newcommand{\geminiIIflashJSSLTestPI }{ 43.70\% }
\newcommand{\geminiIIflashJSSLTestPIII }{ 53.02\% }
\newcommand{\geminiIIflashJSSLTestPV }{ 55.94\% }
\newcommand{\geminiIproJSSLTestPI }{ 50.37\% }
\newcommand{\geminiIproJSSLTestPIII }{ 53.25\% }
\newcommand{\geminiIproJSSLTestPV }{ 54.32\% }
\newcommand{\oIIIminiJSSLTestPI }{ 66.91\% }
\newcommand{\oIIIminiJSSLTestPIII }{ 78.67\% }
\newcommand{\oIIIminiJSSLTestPV }{ 82.70\% }
\newcommand{\claudeJSSHTestPI }{ 41.33\% }
\newcommand{\claudeJSSHTestPIII }{ 54.02\% }
\newcommand{\claudeJSSHTestPV }{ 60.43\% }
\newcommand{\qwenJSSHTestPI }{ 14.44\% }
\newcommand{\qwenJSSHTestPIII }{ 22.02\% }
\newcommand{\qwenJSSHTestPV }{ 26.84\% }
\newcommand{\RIJSSHTestPI }{ 27.59\% }
\newcommand{\RIJSSHTestPIII }{ 44.86\% }
\newcommand{\RIJSSHTestPV }{ 52.85\% }
\newcommand{\geminiIIflashJSSHTestPI }{ 22.80\% }
\newcommand{\geminiIIflashJSSHTestPIII }{ 31.22\% }
\newcommand{\geminiIIflashJSSHTestPV }{ 35.56\% }
\newcommand{\geminiIproJSSHTestPI }{ 23.01\% }
\newcommand{\geminiIproJSSHTestPIII }{ 27.26\% }
\newcommand{\geminiIproJSSHTestPV }{ 29.72\% }
\newcommand{\oIIIminiJSSHTestPI }{ 49.33\% }
\newcommand{\oIIIminiJSSHTestPIII }{ 63.09\% }
\newcommand{\oIIIminiJSSHTestPV }{ 68.56\% }
\newcommand{\claudeJSMHTestPI }{ 35.96\% }
\newcommand{\claudeJSMHTestPIII }{ 43.70\% }
\newcommand{\claudeJSMHTestPV }{ 47.64\% }
\newcommand{\qwenJSMHTestPI }{ 16.84\% }
\newcommand{\qwenJSMHTestPIII }{ 22.11\% }
\newcommand{\qwenJSMHTestPV }{ 24.49\% }
\newcommand{\RIJSMHTestPI }{ 27.62\% }
\newcommand{\RIJSMHTestPIII }{ 40.45\% }
\newcommand{\RIJSMHTestPV }{ 46.13\% }
\newcommand{\geminiIIflashJSMHTestPI }{ 22.11\% }
\newcommand{\geminiIIflashJSMHTestPIII }{ 28.05\% }
\newcommand{\geminiIIflashJSMHTestPV }{ 29.89\% }
\newcommand{\geminiIproJSMHTestPI }{ 12.11\% }
\newcommand{\geminiIproJSMHTestPIII }{ 17.57\% }
\newcommand{\geminiIproJSMHTestPV }{ 19.04\% }
\newcommand{\oIIIminiJSMHTestPI }{ 42.11\% }
\newcommand{\oIIIminiJSMHTestPIII }{ 53.78\% }
\newcommand{\oIIIminiJSMHTestPV }{ 57.10\% }
\newcommand{\cLlamaLJSallSHBASICPI }{ 4.87\% }
\newcommand{\cLlamaLJSallSHBASICPV }{ 14.31\% }

%% file: values_php.tex
\newcommand{\claudePHPallSHBASICPI }{ 13.60\% }
\newcommand{\claudePHPallSHBASICPIII }{ 20.15\% }
\newcommand{\claudePHPallSHBASICPV }{ 23.28\% }
\newcommand{\claudeHaikuPHPallSHBASICPI }{ 13.44\% }
\newcommand{\claudeHaikuPHPallSHBASICPIII }{ 17.74\% }
\newcommand{\claudeHaikuPHPallSHBASICPV }{ 19.64\% }
\newcommand{\gemmaPHPallSHBASICPI }{ 6.37\% }
\newcommand{\gemmaPHPallSHBASICPIII }{ 13.77\% }
\newcommand{\gemmaPHPallSHBASICPV }{ 18.40\% }
\newcommand{\cLlamaSPHPallSHBASICPI }{ 4.28\% }
\newcommand{\cLlamaSPHPallSHBASICPIII }{ 10.93\% }
\newcommand{\cLlamaSPHPallSHBASICPV }{ 15.98\% }
\newcommand{\LlamaIIIPHPallSHBASICPI }{ 9.28\% }
\newcommand{\LlamaIIIPHPallSHBASICPIII }{ 14.46\% }
\newcommand{\LlamaIIIPHPallSHBASICPV }{ 16.48\% }
\newcommand{\CLlamaLPHPallSHBASICPI }{ 2.41\% }
\newcommand{\CLlamaLPHPallSHBASICPIII }{ 6.66\% }
\newcommand{\CLlamaLPHPallSHBASICPV }{ 10.38\% }
\newcommand{\DSCoderPHPallSHBASICPI }{ 6.93\% }
\newcommand{\DSCoderPHPallSHBASICPIII }{ 13.33\% }
\newcommand{\DSCoderPHPallSHBASICPV }{ 17.28\% }
\newcommand{\gptIVoPHPallSHBASICPI }{ 14.55\% }
\newcommand{\gptIVoPHPallSHBASICPIII }{ 20.98\% }
\newcommand{\gptIVoPHPallSHBASICPV }{ 23.96\% }
\newcommand{\qwenPHPallSHBASICPI }{ 10.89\% }
\newcommand{\qwenPHPallSHBASICPIII }{ 13.68\% }
\newcommand{\qwenPHPallSHBASICPV }{ 14.95\% }
\newcommand{\RIPHPallSHBASICPI }{ 18.43\% }
\newcommand{\RIPHPallSHBASICPIII }{ 28.81\% }
\newcommand{\RIPHPallSHBASICPV }{ 33.43\% }
\newcommand{\geminiIIflashPHPallSHBASICPI }{ 14.39\% }
\newcommand{\geminiIIflashPHPallSHBASICPIII }{ 19.61\% }
\newcommand{\geminiIIflashPHPallSHBASICPV }{ 22.06\% }
\newcommand{\geminiIproPHPallSHBASICPI }{ 17.44\% }
\newcommand{\geminiIproPHPallSHBASICPIII }{ 24.05\% }
\newcommand{\geminiIproPHPallSHBASICPV }{ 26.92\% }
\newcommand{\oIIIminiPHPallSHBASICPI }{ 12.63\% }
\newcommand{\oIIIminiPHPallSHBASICPIII }{ 16.20\% }
\newcommand{\oIIIminiPHPallSHBASICPV }{ 17.56\% }
\newcommand{\claudePHPBASICPI }{ 14.08\% }
\newcommand{\claudePHPBASICPIII }{ 19.51\% }
\newcommand{\claudePHPBASICPV }{ 21.65\% }
\newcommand{\qwenPHPBASICPI }{ 12.78\% }
\newcommand{\qwenPHPBASICPIII }{ 15.48\% }
\newcommand{\qwenPHPBASICPV }{ 16.04\% }
\newcommand{\RIPHPBASICPI }{ 17.50\% }
\newcommand{\RIPHPBASICPIII }{ 25.69\% }
\newcommand{\RIPHPBASICPV }{ 29.33\% }
\newcommand{\geminiIIflashPHPBASICPI }{ 17.87\% }
\newcommand{\geminiIIflashPHPBASICPIII }{ 22.84\% }
\newcommand{\geminiIIflashPHPBASICPV }{ 24.39\% }
\newcommand{\geminiIproPHPBASICPI }{ 21.11\% }
\newcommand{\geminiIproPHPBASICPIII }{ 26.93\% }
\newcommand{\geminiIproPHPBASICPV }{ 29.13\% }
\newcommand{\oIIIminiPHPBASICPI }{ 13.33\% }
\newcommand{\oIIIminiPHPBASICPIII }{ 15.60\% }
\newcommand{\oIIIminiPHPBASICPV }{ 16.49\% }
\newcommand{\claudePHPMHBASICPI }{ 0.00\% }
\newcommand{\claudePHPMHBASICPIII }{ 0.00\% }
\newcommand{\claudePHPMHBASICPV }{ 0.00\% }
\newcommand{\oIIIminiPHPMHBASICPI }{ 6.03\% }
\newcommand{\oIIIminiPHPMHBASICPIII }{ 7.80\% }
\newcommand{\oIIIminiPHPMHBASICPV }{ 8.79\% }
\newcommand{\claudePHPSHBASICPI }{ 10.00\% }
\newcommand{\claudePHPSHBASICPIII }{ 12.53\% }
\newcommand{\claudePHPSHBASICPV }{ 13.80\% }
\newcommand{\qwenPHPSHBASICPI }{ 16.46\% }
\newcommand{\qwenPHPSHBASICPIII }{ 21.17\% }
\newcommand{\qwenPHPSHBASICPV }{ 21.79\% }
\newcommand{\RIPHPSHBASICPI }{ 14.67\% }
\newcommand{\RIPHPSHBASICPIII }{ 24.09\% }
\newcommand{\RIPHPSHBASICPV }{ 29.44\% }
\newcommand{\geminiIIflashPHPSHBASICPI }{ 17.50\% }
\newcommand{\geminiIIflashPHPSHBASICPIII }{ 20.75\% }
\newcommand{\geminiIIflashPHPSHBASICPV }{ 22.07\% }
\newcommand{\geminiIproPHPSHBASICPI }{ 17.50\% }
\newcommand{\geminiIproPHPSHBASICPIII }{ 21.55\% }
\newcommand{\geminiIproPHPSHBASICPV }{ 24.14\% }
\newcommand{\oIIIminiPHPSHBASICPI }{ 8.33\% }
\newcommand{\oIIIminiPHPSHBASICPIII }{ 9.86\% }
\newcommand{\oIIIminiPHPSHBASICPV }{ 10.41\% }
\newcommand{\claudePHPSLBASICPI }{ 20.61\% }
\newcommand{\claudePHPSLBASICPIII }{ 29.84\% }
\newcommand{\claudePHPSLBASICPV }{ 33.19\% }
\newcommand{\qwenPHPSLBASICPI }{ 8.95\% }
\newcommand{\qwenPHPSLBASICPIII }{ 9.63\% }
\newcommand{\qwenPHPSLBASICPV }{ 10.20\% }
\newcommand{\RIPHPSLBASICPI }{ 21.25\% }
\newcommand{\RIPHPSLBASICPIII }{ 28.80\% }
\newcommand{\RIPHPSLBASICPV }{ 31.05\% }
\newcommand{\geminiIIflashPHPSLBASICPI }{ 19.05\% }
\newcommand{\geminiIIflashPHPSLBASICPIII }{ 25.29\% }
\newcommand{\geminiIIflashPHPSLBASICPV }{ 27.14\% }
\newcommand{\geminiIproPHPSLBASICPI }{ 24.95\% }
\newcommand{\geminiIproPHPSLBASICPIII }{ 32.84\% }
\newcommand{\geminiIproPHPSLBASICPV }{ 35.00\% }
\newcommand{\oIIIminiPHPSLBASICPI }{ 21.33\% }
\newcommand{\oIIIminiPHPSLBASICPIII }{ 24.89\% }
\newcommand{\oIIIminiPHPSLBASICPV }{ 26.17\% }

\newcommand{\claudePHPPerfectLLPI }{ 14.91\% }
\newcommand{\claudePHPPerfectLLPIII }{ 20.27\% }
\newcommand{\claudePHPPerfectLLPV }{ 22.52\% }
\newcommand{\qwenPHPPerfectLLPI }{ 10.82\% }
\newcommand{\qwenPHPPerfectLLPIII }{ 15.37\% }
\newcommand{\qwenPHPPerfectLLPV }{ 17.52\% }
\newcommand{\RIPHPPerfectLLPI }{ 17.39\% }
\newcommand{\RIPHPPerfectLLPIII }{ 24.97\% }
\newcommand{\RIPHPPerfectLLPV }{ 28.67\% }
\newcommand{\geminiIIflashPHPPerfectLLPI }{ 13.75\% }
\newcommand{\geminiIIflashPHPPerfectLLPIII }{ 16.57\% }
\newcommand{\geminiIIflashPHPPerfectLLPV }{ 17.92\% }
\newcommand{\geminiIproPHPPerfectLLPI }{ 14.36\% }
\newcommand{\geminiIproPHPPerfectLLPIII }{ 19.46\% }
\newcommand{\geminiIproPHPPerfectLLPV }{ 21.86\% }
\newcommand{\oIIIminiPHPPerfectLLPI }{ 16.01\% }
\newcommand{\oIIIminiPHPPerfectLLPIII }{ 20.66\% }
\newcommand{\oIIIminiPHPPerfectLLPV }{ 22.68\% }
\newcommand{\claudePHPMHPerfectLLPI }{ 6.41\% }
\newcommand{\claudePHPMHPerfectLLPIII }{ 12.34\% }
\newcommand{\claudePHPMHPerfectLLPV }{ 14.99\% }
\newcommand{\qwenPHPMHPerfectLLPI }{ 2.05\% }
\newcommand{\qwenPHPMHPerfectLLPIII }{ 4.56\% }
\newcommand{\qwenPHPMHPerfectLLPV }{ 5.88\% }
\newcommand{\RIPHPMHPerfectLLPI }{ 11.54\% }
\newcommand{\RIPHPMHPerfectLLPIII }{ 14.74\% }
\newcommand{\RIPHPMHPerfectLLPV }{ 15.31\% }
\newcommand{\geminiIIflashPHPMHPerfectLLPI }{ 5.64\% }
\newcommand{\geminiIIflashPHPMHPerfectLLPIII }{ 8.66\% }
\newcommand{\geminiIIflashPHPMHPerfectLLPV }{ 10.20\% }
\newcommand{\geminiIproPHPMHPerfectLLPI }{ 2.05\% }
\newcommand{\geminiIproPHPMHPerfectLLPIII }{ 5.42\% }
\newcommand{\geminiIproPHPMHPerfectLLPV }{ 8.02\% }
\newcommand{\oIIIminiPHPMHPerfectLLPI }{ 7.18\% }
\newcommand{\oIIIminiPHPMHPerfectLLPIII }{ 7.69\% }
\newcommand{\oIIIminiPHPMHPerfectLLPV }{ 7.69\% }
\newcommand{\claudePHPSHPerfectLLPI }{ 12.50\% }
\newcommand{\claudePHPSHPerfectLLPIII }{ 16.59\% }
\newcommand{\claudePHPSHPerfectLLPV }{ 18.48\% }
\newcommand{\qwenPHPSHPerfectLLPI }{ 15.62\% }
\newcommand{\qwenPHPSHPerfectLLPIII }{ 20.74\% }
\newcommand{\qwenPHPSHPerfectLLPV }{ 22.70\% }
\newcommand{\RIPHPSHPerfectLLPI }{ 15.00\% }
\newcommand{\RIPHPSHPerfectLLPIII }{ 21.83\% }
\newcommand{\RIPHPSHPerfectLLPV }{ 25.45\% }
\newcommand{\geminiIIflashPHPSHPerfectLLPI }{ 17.08\% }
\newcommand{\geminiIIflashPHPSHPerfectLLPIII }{ 19.33\% }
\newcommand{\geminiIIflashPHPSHPerfectLLPV }{ 20.40\% }
\newcommand{\geminiIproPHPSHPerfectLLPI }{ 19.37\% }
\newcommand{\geminiIproPHPSHPerfectLLPIII }{ 23.48\% }
\newcommand{\geminiIproPHPSHPerfectLLPV }{ 24.49\% }
\newcommand{\oIIIminiPHPSHPerfectLLPI }{ 10.11\% }
\newcommand{\oIIIminiPHPSHPerfectLLPIII }{ 16.91\% }
\newcommand{\oIIIminiPHPSHPerfectLLPV }{ 20.28\% }
\newcommand{\claudePHPSLPerfectLLPI }{ 23.43\% }
\newcommand{\claudePHPSLPerfectLLPIII }{ 29.51\% }
\newcommand{\claudePHPSLPerfectLLPV }{ 31.81\% }
\newcommand{\qwenPHPSLPerfectLLPI }{ 12.95\% }
\newcommand{\qwenPHPSLPerfectLLPIII }{ 18.48\% }
\newcommand{\qwenPHPSLPerfectLLPV }{ 21.42\% }
\newcommand{\RIPHPSLPerfectLLPI }{ 24.12\% }
\newcommand{\RIPHPSLPerfectLLPIII }{ 35.75\% }
\newcommand{\RIPHPSLPerfectLLPV }{ 41.91\% }
\newcommand{\geminiIIflashPHPSLPerfectLLPI }{ 17.90\% }
\newcommand{\geminiIIflashPHPSLPerfectLLPIII }{ 21.33\% }
\newcommand{\geminiIIflashPHPSLPerfectLLPV }{ 22.93\% }
\newcommand{\geminiIproPHPSLPerfectLLPI }{ 20.57\% }
\newcommand{\geminiIproPHPSLPerfectLLPIII }{ 28.43\% }
\newcommand{\geminiIproPHPSLPerfectLLPV }{ 32.23\% }
\newcommand{\oIIIminiPHPSLPerfectLLPI }{ 27.81\% }
\newcommand{\oIIIminiPHPSLPerfectLLPIII }{ 33.63\% }
\newcommand{\oIIIminiPHPSLPerfectLLPV }{ 35.94\% }
\newcommand{\claudePHPTestPI }{ 39.06\% }
\newcommand{\claudePHPTestPIII }{ 47.83\% }
\newcommand{\claudePHPTestPV }{ 50.25\% }
\newcommand{\qwenPHPTestPI }{ 28.24\% }
\newcommand{\qwenPHPTestPIII }{ 36.79\% }
\newcommand{\qwenPHPTestPV }{ 40.00\% }
\newcommand{\RIPHPTestPI }{ 38.97\% }
\newcommand{\RIPHPTestPIII }{ 52.14\% }
\newcommand{\RIPHPTestPV }{ 57.35\% }
\newcommand{\geminiIIflashPHPTestPI }{ 27.96\% }
\newcommand{\geminiIIflashPHPTestPIII }{ 34.38\% }
\newcommand{\geminiIIflashPHPTestPV }{ 37.52\% }
\newcommand{\geminiIproPHPTestPI }{ 31.76\% }
\newcommand{\geminiIproPHPTestPIII }{ 38.94\% }
\newcommand{\geminiIproPHPTestPV }{ 41.78\% }
\newcommand{\oIIIminiPHPTestPI }{ 45.04\% }
\newcommand{\oIIIminiPHPTestPIII }{ 54.71\% }
\newcommand{\oIIIminiPHPTestPV }{ 58.44\% }
\newcommand{\claudePHPMHTestPI }{ 25.90\% }
\newcommand{\claudePHPMHTestPIII }{ 32.02\% }
\newcommand{\claudePHPMHTestPV }{ 34.56\% }
\newcommand{\qwenPHPMHTestPI }{ 19.23\% }
\newcommand{\qwenPHPMHTestPIII }{ 26.91\% }
\newcommand{\qwenPHPMHTestPV }{ 29.29\% }
\newcommand{\RIPHPMHTestPI }{ 28.72\% }
\newcommand{\RIPHPMHTestPIII }{ 44.32\% }
\newcommand{\RIPHPMHTestPV }{ 50.83\% }
\newcommand{\geminiIIflashPHPMHTestPI }{ 12.05\% }
\newcommand{\geminiIIflashPHPMHTestPIII }{ 17.33\% }
\newcommand{\geminiIIflashPHPMHTestPV }{ 20.70\% }
\newcommand{\geminiIproPHPMHTestPI }{ 25.64\% }
\newcommand{\geminiIproPHPMHTestPIII }{ 28.32\% }
\newcommand{\geminiIproPHPMHTestPV }{ 29.12\% }
\newcommand{\oIIIminiPHPMHTestPI }{ 28.72\% }
\newcommand{\oIIIminiPHPMHTestPIII }{ 34.58\% }
\newcommand{\oIIIminiPHPMHTestPV }{ 37.55\% }
\newcommand{\claudePHPSHTestPI }{ 45.42\% }
\newcommand{\claudePHPSHTestPIII }{ 52.24\% }
\newcommand{\claudePHPSHTestPV }{ 53.04\% }
\newcommand{\qwenPHPSHTestPI }{ 35.21\% }
\newcommand{\qwenPHPSHTestPIII }{ 40.84\% }
\newcommand{\qwenPHPSHTestPV }{ 42.44\% }
\newcommand{\RIPHPSHTestPI }{ 42.15\% }
\newcommand{\RIPHPSHTestPIII }{ 54.43\% }
\newcommand{\RIPHPSHTestPV }{ 59.09\% }
\newcommand{\geminiIIflashPHPSHTestPI }{ 31.87\% }
\newcommand{\geminiIIflashPHPSHTestPIII }{ 41.07\% }
\newcommand{\geminiIIflashPHPSHTestPV }{ 43.84\% }
\newcommand{\geminiIproPHPSHTestPI }{ 30.21\% }
\newcommand{\geminiIproPHPSHTestPIII }{ 36.83\% }
\newcommand{\geminiIproPHPSHTestPV }{ 39.75\% }
\newcommand{\oIIIminiPHPSHTestPI }{ 44.30\% }
\newcommand{\oIIIminiPHPSHTestPIII }{ 53.69\% }
\newcommand{\oIIIminiPHPSHTestPV }{ 57.26\% }
\newcommand{\claudePHPSLTestPI }{ 43.14\% }
\newcommand{\claudePHPSLTestPIII }{ 55.77\% }
\newcommand{\claudePHPSLTestPV }{ 59.64\% }
\newcommand{\qwenPHPSLTestPI }{ 28.57\% }
\newcommand{\qwenPHPSLTestPIII }{ 40.42\% }
\newcommand{\qwenPHPSLTestPV }{ 45.73\% }
\newcommand{\RIPHPSLTestPI }{ 43.92\% }
\newcommand{\RIPHPSLTestPIII }{ 56.02\% }
\newcommand{\RIPHPSLTestPV }{ 60.76\% }
\newcommand{\geminiIIflashPHPSLTestPI }{ 36.19\% }
\newcommand{\geminiIIflashPHPSLTestPIII }{ 40.93\% }
\newcommand{\geminiIIflashPHPSLTestPV }{ 44.25\% }
\newcommand{\geminiIproPHPSLTestPI }{ 37.71\% }
\newcommand{\geminiIproPHPSLTestPIII }{ 48.77\% }
\newcommand{\geminiIproPHPSLTestPV }{ 53.05\% }
\newcommand{\oIIIminiPHPSLTestPI }{ 58.59\% }
\newcommand{\oIIIminiPHPSLTestPIII }{ 71.52\% }
\newcommand{\oIIIminiPHPSLTestPV }{ 76.00\% }

%% file: values_python.tex
\newcommand{\claudePythonallSHBASICPI }{ 7.59\% }
\newcommand{\claudePythonallSHBASICPIII }{ 11.63\% }
\newcommand{\claudePythonallSHBASICPV }{ 13.64\% }
\newcommand{\claudeHaikuPythonallSHBASICPI }{ 7.28\% }
\newcommand{\claudeHaikuPythonallSHBASICPIII }{ 9.57\% }
\newcommand{\claudeHaikuPythonallSHBASICPV }{ 10.49\% }
\newcommand{\gemmaPythonallSHBASICPI }{ 3.49\% }
\newcommand{\gemmaPythonallSHBASICPIII }{ 7.26\% }
\newcommand{\gemmaPythonallSHBASICPV }{ 9.48\% }
\newcommand{\cLlamaSPythonallSHBASICPI }{ 1.95\% }
\newcommand{\cLlamaSPythonallSHBASICPIII }{ 5.14\% }
\newcommand{\cLlamaSPythonallSHBASICPV }{ 7.63\% }
\newcommand{\LlamaIIIPythonallSHBASICPI }{ 5.33\% }
\newcommand{\LlamaIIIPythonallSHBASICPIII }{ 7.92\% }
\newcommand{\LlamaIIIPythonallSHBASICPV }{ 9.29\% }
\newcommand{\CLlamaLPythonallSHBASICPI }{ 1.13\% }
\newcommand{\CLlamaLPythonallSHBASICPIII }{ 2.58\% }
\newcommand{\CLlamaLPythonallSHBASICPV }{ 3.56\% }
\newcommand{\DSCoderPythonallSHBASICPI }{ 3.51\% }
\newcommand{\DSCoderPythonallSHBASICPIII }{ 6.90\% }
\newcommand{\DSCoderPythonallSHBASICPV }{ 9.19\% }
\newcommand{\gptIVoPythonallSHBASICPI }{ 3.94\% }
\newcommand{\gptIVoPythonallSHBASICPIII }{ 9.26\% }
\newcommand{\gptIVoPythonallSHBASICPV }{ 12.56\% }
\newcommand{\qwenPythonallSHBASICPI }{ 3.79\% }
\newcommand{\qwenPythonallSHBASICPIII }{ 5.29\% }
\newcommand{\qwenPythonallSHBASICPV }{ 6.43\% }
\newcommand{\RIPythonallSHBASICPI }{ 6.56\% }
\newcommand{\RIPythonallSHBASICPIII }{ 12.11\% }
\newcommand{\RIPythonallSHBASICPV }{ 15.63\% }
\newcommand{\geminiIIflashPythonallSHBASICPI }{ 12.51\% }
\newcommand{\geminiIIflashPythonallSHBASICPIII }{ 16.16\% }
\newcommand{\geminiIIflashPythonallSHBASICPV }{ 18.03\% }
\newcommand{\geminiIproPythonallSHBASICPI }{ 6.36\% }
\newcommand{\geminiIproPythonallSHBASICPIII }{ 11.25\% }
\newcommand{\geminiIproPythonallSHBASICPV }{ 13.30\% }
\newcommand{\oIIIminiPythonallSHBASICPI }{ 8.92\% }
\newcommand{\oIIIminiPythonallSHBASICPIII }{ 10.67\% }
\newcommand{\oIIIminiPythonallSHBASICPV }{ 11.50\% }
\newcommand{\claudePythonBASICPI }{ 5.91\% }
\newcommand{\claudePythonBASICPIII }{ 9.20\% }
\newcommand{\claudePythonBASICPV }{ 10.71\% }
\newcommand{\qwenPythonBASICPI }{ 2.72\% }
\newcommand{\qwenPythonBASICPIII }{ 3.91\% }
\newcommand{\qwenPythonBASICPV }{ 4.85\% }
\newcommand{\RIPythonBASICPI }{ 6.27\% }
\newcommand{\RIPythonBASICPIII }{ 11.57\% }
\newcommand{\RIPythonBASICPV }{ 14.94\% }
\newcommand{\geminiIIflashPythonBASICPI }{ 11.35\% }
\newcommand{\geminiIIflashPythonBASICPIII }{ 14.74\% }
\newcommand{\geminiIIflashPythonBASICPV }{ 16.27\% }
\newcommand{\geminiIproPythonBASICPI }{ 6.67\% }
\newcommand{\geminiIproPythonBASICPIII }{ 10.62\% }
\newcommand{\geminiIproPythonBASICPV }{ 12.12\% }
\newcommand{\oIIIminiPythonBASICPI }{ 6.24\% }
\newcommand{\oIIIminiPythonBASICPIII }{ 7.46\% }
\newcommand{\oIIIminiPythonBASICPV }{ 8.04\% }
\newcommand{\claudePythonSLBASICPI }{ 2.54\% }
\newcommand{\claudePythonSLBASICPIII }{ 5.97\% }
\newcommand{\claudePythonSLBASICPV }{ 7.87\% }
\newcommand{\qwenPythonSLBASICPI }{ 0.63\% }
\newcommand{\qwenPythonSLBASICPIII }{ 1.90\% }
\newcommand{\qwenPythonSLBASICPV }{ 3.17\% }
\newcommand{\RIPythonSLBASICPI }{ 4.44\% }
\newcommand{\RIPythonSLBASICPIII }{ 8.19\% }
\newcommand{\RIPythonSLBASICPV }{ 10.20\% }
\newcommand{\geminiIIflashPythonSLBASICPI }{ 15.24\% }
\newcommand{\geminiIIflashPythonSLBASICPIII }{ 16.75\% }
\newcommand{\geminiIIflashPythonSLBASICPV }{ 17.79\% }
\newcommand{\geminiIproPythonSLBASICPI }{ 6.67\% }
\newcommand{\geminiIproPythonSLBASICPIII }{ 13.22\% }
\newcommand{\geminiIproPythonSLBASICPV }{ 16.45\% }
\newcommand{\oIIIminiPythonSLBASICPI }{ 4.44\% }
\newcommand{\oIIIminiPythonSLBASICPIII }{ 6.52\% }
\newcommand{\oIIIminiPythonSLBASICPV }{ 7.48\% }
\newcommand{\claudePythonSHBASICPI }{ 10.73\% }
\newcommand{\claudePythonSHBASICPIII }{ 15.38\% }
\newcommand{\claudePythonSHBASICPV }{ 17.59\% }
\newcommand{\qwenPythonSHBASICPI }{ 5.69\% }
\newcommand{\qwenPythonSHBASICPIII }{ 7.41\% }
\newcommand{\qwenPythonSHBASICPV }{ 8.57\% }
\newcommand{\RIPythonSHBASICPI }{ 7.64\% }
\newcommand{\RIPythonSHBASICPIII }{ 13.74\% }
\newcommand{\RIPythonSHBASICPV }{ 17.77\% }
\newcommand{\geminiIIflashPythonSHBASICPI }{ 12.03\% }
\newcommand{\geminiIIflashPythonSHBASICPIII }{ 17.05\% }
\newcommand{\geminiIIflashPythonSHBASICPV }{ 19.48\% }
\newcommand{\geminiIproPythonSHBASICPI }{ 6.67\% }
\newcommand{\geminiIproPythonSHBASICPIII }{ 11.07\% }
\newcommand{\geminiIproPythonSHBASICPV }{ 12.68\% }
\newcommand{\oIIIminiPythonSHBASICPI }{ 11.87\% }
\newcommand{\oIIIminiPythonSHBASICPIII }{ 13.58\% }
\newcommand{\oIIIminiPythonSHBASICPV }{ 14.40\% }
\newcommand{\claudePythonMHBASICPI }{ 1.67\% }
\newcommand{\claudePythonMHBASICPIII }{ 3.00\% }
\newcommand{\claudePythonMHBASICPV }{ 3.29\% }
\newcommand{\qwenPythonMHBASICPI }{ 0.22\% }
\newcommand{\qwenPythonMHBASICPIII }{ 0.65\% }
\newcommand{\qwenPythonMHBASICPV }{ 1.08\% }
\newcommand{\RIPythonMHBASICPI }{ 3.33\% }
\newcommand{\RIPythonMHBASICPIII }{ 8.61\% }
\newcommand{\RIPythonMHBASICPV }{ 12.27\% }
\newcommand{\geminiIIflashPythonMHBASICPI }{ 7.92\% }
\newcommand{\geminiIIflashPythonMHBASICPIII }{ 10.46\% }
\newcommand{\geminiIIflashPythonMHBASICPV }{ 11.16\% }
\newcommand{\geminiIproPythonMHBASICPI }{ 6.67\% }
\newcommand{\geminiIproPythonMHBASICPIII }{ 8.46\% }
\newcommand{\geminiIproPythonMHBASICPV }{ 8.77\% }
\newcommand{\oIIIminiPythonMHBASICPI }{ 0.00\% }
\newcommand{\oIIIminiPythonMHBASICPIII }{ 0.00\% }
\newcommand{\oIIIminiPythonMHBASICPV }{ 0.00\% }
\newcommand{\claudePythonPerfectLLPI }{ 8.89\% }
\newcommand{\claudePythonPerfectLLPIII }{ 13.23\% }
\newcommand{\claudePythonPerfectLLPV }{ 15.18\% }
\newcommand{\qwenPythonPerfectLLPI }{ 3.58\% }
\newcommand{\qwenPythonPerfectLLPIII }{ 5.10\% }
\newcommand{\qwenPythonPerfectLLPV }{ 5.95\% }
\newcommand{\RIPythonPerfectLLPI }{ 6.46\% }
\newcommand{\RIPythonPerfectLLPIII }{ 11.90\% }
\newcommand{\RIPythonPerfectLLPV }{ 14.91\% }
\newcommand{\geminiIIflashPythonPerfectLLPI }{ 13.62\% }
\newcommand{\geminiIIflashPythonPerfectLLPIII }{ 18.92\% }
\newcommand{\geminiIIflashPythonPerfectLLPV }{ 21.43\% }
\newcommand{\geminiIproPythonPerfectLLPI }{ 9.36\% }
\newcommand{\geminiIproPythonPerfectLLPIII }{ 14.70\% }
\newcommand{\geminiIproPythonPerfectLLPV }{ 16.74\% }
\newcommand{\oIIIminiPythonPerfectLLPI }{ 9.12\% }
\newcommand{\oIIIminiPythonPerfectLLPIII }{ 11.37\% }
\newcommand{\oIIIminiPythonPerfectLLPV }{ 12.60\% }
\newcommand{\claudePythonSLPerfectLLPI }{ 4.13\% }
\newcommand{\claudePythonSLPerfectLLPIII }{ 7.59\% }
\newcommand{\claudePythonSLPerfectLLPV }{ 8.78\% }
\newcommand{\qwenPythonSLPerfectLLPI }{ 6.03\% }
\newcommand{\qwenPythonSLPerfectLLPIII }{ 10.69\% }
\newcommand{\qwenPythonSLPerfectLLPV }{ 12.63\% }
\newcommand{\RIPythonSLPerfectLLPI }{ 2.22\% }
\newcommand{\RIPythonSLPerfectLLPIII }{ 6.39\% }
\newcommand{\RIPythonSLPerfectLLPV }{ 10.20\% }
\newcommand{\geminiIIflashPythonSLPerfectLLPI }{ 19.05\% }
\newcommand{\geminiIIflashPythonSLPerfectLLPIII }{ 26.56\% }
\newcommand{\geminiIIflashPythonSLPerfectLLPV }{ 29.75\% }
\newcommand{\geminiIproPythonSLPerfectLLPI }{ 4.76\% }
\newcommand{\geminiIproPythonSLPerfectLLPIII }{ 10.23\% }
\newcommand{\geminiIproPythonSLPerfectLLPV }{ 12.63\% }
\newcommand{\oIIIminiPythonSLPerfectLLPI }{ 8.25\% }
\newcommand{\oIIIminiPythonSLPerfectLLPIII }{ 11.62\% }
\newcommand{\oIIIminiPythonSLPerfectLLPV }{ 13.00\% }
\newcommand{\claudePythonSHPerfectLLPI }{ 14.96\% }
\newcommand{\claudePythonSHPerfectLLPIII }{ 21.11\% }
\newcommand{\claudePythonSHPerfectLLPV }{ 24.00\% }
\newcommand{\qwenPythonSHPerfectLLPI }{ 4.88\% }
\newcommand{\qwenPythonSHPerfectLLPIII }{ 5.37\% }
\newcommand{\qwenPythonSHPerfectLLPV }{ 5.69\% }
\newcommand{\RIPythonSHPerfectLLPI }{ 11.87\% }
\newcommand{\RIPythonSHPerfectLLPIII }{ 19.73\% }
\newcommand{\RIPythonSHPerfectLLPV }{ 23.05\% }
\newcommand{\geminiIIflashPythonSHPerfectLLPI }{ 12.03\% }
\newcommand{\geminiIIflashPythonSHPerfectLLPIII }{ 14.32\% }
\newcommand{\geminiIIflashPythonSHPerfectLLPV }{ 15.61\% }
\newcommand{\geminiIproPythonSHPerfectLLPI }{ 9.76\% }
\newcommand{\geminiIproPythonSHPerfectLLPIII }{ 12.77\% }
\newcommand{\geminiIproPythonSHPerfectLLPV }{ 14.26\% }
\newcommand{\oIIIminiPythonSHPerfectLLPI }{ 14.80\% }
\newcommand{\oIIIminiPythonSHPerfectLLPIII }{ 17.00\% }
\newcommand{\oIIIminiPythonSHPerfectLLPV }{ 18.47\% }
\newcommand{\claudePythonMHPerfectLLPI }{ 4.51\% }
\newcommand{\claudePythonMHPerfectLLPIII }{ 7.20\% }
\newcommand{\claudePythonMHPerfectLLPV }{ 8.49\% }
\newcommand{\qwenPythonMHPerfectLLPI }{ 0.40\% }
\newcommand{\qwenPythonMHPerfectLLPIII }{ 1.21\% }
\newcommand{\qwenPythonMHPerfectLLPV }{ 2.02\% }
\newcommand{\RIPythonMHPerfectLLPI }{ 2.55\% }
\newcommand{\RIPythonMHPerfectLLPIII }{ 5.86\% }
\newcommand{\RIPythonMHPerfectLLPV }{ 7.99\% }
\newcommand{\geminiIIflashPythonMHPerfectLLPI }{ 12.08\% }
\newcommand{\geminiIIflashPythonMHPerfectLLPIII }{ 19.81\% }
\newcommand{\geminiIIflashPythonMHPerfectLLPV }{ 23.43\% }
\newcommand{\geminiIproPythonMHPerfectLLPI }{ 11.88\% }
\newcommand{\geminiIproPythonMHPerfectLLPIII }{ 20.11\% }
\newcommand{\geminiIproPythonMHPerfectLLPV }{ 22.62\% }
\newcommand{\oIIIminiPythonMHPerfectLLPI }{ 2.63\% }
\newcommand{\oIIIminiPythonMHPerfectLLPIII }{ 4.22\% }
\newcommand{\oIIIminiPythonMHPerfectLLPV }{ 5.05\% }
\newcommand{\claudePythonTestPI }{ 29.17\% }
\newcommand{\claudePythonTestPIII }{ 34.51\% }
\newcommand{\claudePythonTestPV }{ 36.76\% }
\newcommand{\qwenPythonTestPI }{ 17.01\% }
\newcommand{\qwenPythonTestPIII }{ 26.15\% }
\newcommand{\qwenPythonTestPV }{ 28.98\% }
\newcommand{\RIPythonTestPI }{ 40.07\% }
\newcommand{\RIPythonTestPIII }{ 58.05\% }
\newcommand{\RIPythonTestPV }{ 63.87\% }
\newcommand{\geminiIIflashPythonTestPI }{ 39.79\% }
\newcommand{\geminiIIflashPythonTestPIII }{ 45.44\% }
\newcommand{\geminiIIflashPythonTestPV }{ 48.14\% }
\newcommand{\geminiIproPythonTestPI }{ 38.89\% }
\newcommand{\geminiIproPythonTestPIII }{ 45.10\% }
\newcommand{\geminiIproPythonTestPV }{ 47.41\% }
\newcommand{\oIIIminiPythonTestPI }{ 35.69\% }
\newcommand{\oIIIminiPythonTestPIII }{ 41.61\% }
\newcommand{\oIIIminiPythonTestPV }{ 43.42\% }
\newcommand{\claudePythonSLTestPI }{ 24.76\% }
\newcommand{\claudePythonSLTestPIII }{ 30.21\% }
\newcommand{\claudePythonSLTestPV }{ 32.93\% }
\newcommand{\qwenPythonSLTestPI }{ 20.63\% }
\newcommand{\qwenPythonSLTestPIII }{ 31.53\% }
\newcommand{\qwenPythonSLTestPV }{ 34.67\% }
\newcommand{\RIPythonSLTestPI }{ 47.94\% }
\newcommand{\RIPythonSLTestPIII }{ 66.28\% }
\newcommand{\RIPythonSLTestPV }{ 73.00\% }
\newcommand{\geminiIIflashPythonSLTestPI }{ 44.44\% }
\newcommand{\geminiIIflashPythonSLTestPIII }{ 48.30\% }
\newcommand{\geminiIIflashPythonSLTestPV }{ 51.47\% }
\newcommand{\geminiIproPythonSLTestPI }{ 63.81\% }
\newcommand{\geminiIproPythonSLTestPIII }{ 69.48\% }
\newcommand{\geminiIproPythonSLTestPV }{ 71.67\% }
\newcommand{\oIIIminiPythonSLTestPI }{ 34.29\% }
\newcommand{\oIIIminiPythonSLTestPIII }{ 41.15\% }
\newcommand{\oIIIminiPythonSLTestPV }{ 43.18\% }
\newcommand{\claudePythonSHTestPI }{ 34.63\% }
\newcommand{\claudePythonSHTestPIII }{ 40.77\% }
\newcommand{\claudePythonSHTestPV }{ 44.03\% }
\newcommand{\qwenPythonSHTestPI }{ 22.28\% }
\newcommand{\qwenPythonSHTestPIII }{ 33.03\% }
\newcommand{\qwenPythonSHTestPV }{ 36.16\% }
\newcommand{\RIPythonSHTestPI }{ 45.20\% }
\newcommand{\RIPythonSHTestPIII }{ 64.96\% }
\newcommand{\RIPythonSHTestPV }{ 71.08\% }
\newcommand{\geminiIIflashPythonSHTestPI }{ 45.20\% }
\newcommand{\geminiIIflashPythonSHTestPIII }{ 51.01\% }
\newcommand{\geminiIIflashPythonSHTestPV }{ 53.21\% }
\newcommand{\geminiIproPythonSHTestPI }{ 35.61\% }
\newcommand{\geminiIproPythonSHTestPIII }{ 42.21\% }
\newcommand{\geminiIproPythonSHTestPV }{ 45.00\% }
\newcommand{\oIIIminiPythonSHTestPI }{ 40.00\% }
\newcommand{\oIIIminiPythonSHTestPIII }{ 44.76\% }
\newcommand{\oIIIminiPythonSHTestPV }{ 46.29\% }
\newcommand{\claudePythonMHTestPI }{ 25.29\% }
\newcommand{\claudePythonMHTestPIII }{ 29.61\% }
\newcommand{\claudePythonMHTestPV }{ 30.37\% }
\newcommand{\qwenPythonMHTestPI }{ 8.43\% }
\newcommand{\qwenPythonMHTestPIII }{ 14.52\% }
\newcommand{\qwenPythonMHTestPV }{ 16.80\% }
\newcommand{\RIPythonMHTestPI }{ 28.69\% }
\newcommand{\RIPythonMHTestPIII }{ 44.22\% }
\newcommand{\RIPythonMHTestPV }{ 49.11\% }
\newcommand{\geminiIIflashPythonMHTestPI }{ 30.39\% }
\newcommand{\geminiIIflashPythonMHTestPIII }{ 36.95\% }
\newcommand{\geminiIIflashPythonMHTestPV }{ 39.97\% }
\newcommand{\geminiIproPythonMHTestPI }{ 27.45\% }
\newcommand{\geminiIproPythonMHTestPIII }{ 33.52\% }
\newcommand{\geminiIproPythonMHTestPV }{ 35.32\% }
\newcommand{\oIIIminiPythonMHTestPI }{ 31.37\% }
\newcommand{\oIIIminiPythonMHTestPIII }{ 38.11\% }
\newcommand{\oIIIminiPythonMHTestPV }{ 40.11\% }

%% file: values.tex
\newcommand{\qwenclaudeJSBASICPV }{ 20.94\% }
\newcommand{\RIclaudeJSBASICPV }{ 21.34\% }
\newcommand{\geminiIIflashclaudeJSBASICPV }{ 17.89\% }
\newcommand{\claudegeminiIproJSBASICPV }{ 21.35\% }
\newcommand{\claudeoIIIminiJSBASICPV }{ 18.92\% }
\newcommand{\qwenRIJSBASICPV }{ 17.69\% }
\newcommand{\qwengeminiIIflashJSBASICPV }{ 17.42\% }
\newcommand{\qwengeminiIproJSBASICPV }{ 21.41\% }
\newcommand{\qwenoIIIminiJSBASICPV }{ 17.50\% }
\newcommand{\geminiIIflashRIJSBASICPV }{ 17.77\% }
\newcommand{\geminiIproRIJSBASICPV }{ 17.77\% }
\newcommand{\RIoIIIminiJSBASICPV }{ 18.94\% }
\newcommand{\geminiIIflashgeminiIproJSBASICPV }{ 17.47\% }
\newcommand{\geminiIIflashoIIIminiJSBASICPV }{ 15.89\% }
\newcommand{\geminiIprooIIIminiJSBASICPV }{ 15.98\% }
\newcommand{\claudeqwenJSPerfectLLPV }{ 28.58\% }
\newcommand{\claudeRIJSPerfectLLPV }{ 35.76\% }
\newcommand{\claudegeminiIIflashJSPerfectLLPV }{ 32.14\% }
\newcommand{\claudegeminiIproJSPerfectLLPV }{ 32.45\% }
\newcommand{\claudeoIIIminiJSPerfectLLPV }{ 33.19\% }
\newcommand{\RIqwenJSPerfectLLPV }{ 28.70\% }
\newcommand{\geminiIIflashqwenJSPerfectLLPV }{ 24.05\% }
\newcommand{\geminiIproqwenJSPerfectLLPV }{ 29.02\% }
\newcommand{\oIIIminiqwenJSPerfectLLPV }{ 29.04\% }
\newcommand{\geminiIIflashRIJSPerfectLLPV }{ 29.64\% }
\newcommand{\geminiIproRIJSPerfectLLPV }{ 35.37\% }
\newcommand{\oIIIminiRIJSPerfectLLPV }{ 32.21\% }
\newcommand{\geminiIprogeminiIIflashJSPerfectLLPV }{ 31.17\% }
\newcommand{\geminiIIflashoIIIminiJSPerfectLLPV }{ 28.27\% }
\newcommand{\geminiIprooIIIminiJSPerfectLLPV }{ 33.55\% }
\newcommand{\claudeqwenJSTestPV }{ 54.19\% }
\newcommand{\claudeRIJSTestPV }{ 63.48\% }
\newcommand{\claudegeminiIIflashJSTestPV }{ 57.11\% }
\newcommand{\claudegeminiIproJSTestPV }{ 55.11\% }
\newcommand{\oIIIminiclaudeJSTestPV }{ 68.77\% }
\newcommand{\RIqwenJSTestPV }{ 55.31\% }
\newcommand{\geminiIIflashqwenJSTestPV }{ 40.80\% }
\newcommand{\geminiIproqwenJSTestPV }{ 39.03\% }
\newcommand{\oIIIminiqwenJSTestPV }{ 66.73\% }
\newcommand{\RIgeminiIIflashJSTestPV }{ 56.23\% }
\newcommand{\RIgeminiIproJSTestPV }{ 56.08\% }
\newcommand{\oIIIminiRIJSTestPV }{ 71.68\% }
\newcommand{\geminiIIflashgeminiIproJSTestPV }{ 43.97\% }
\newcommand{\oIIIminigeminiIIflashJSTestPV }{ 68.50\% }
\newcommand{\oIIIminigeminiIproJSTestPV }{ 66.33\% }

\newcommand{\claudeqwenJavaPerfectLLPV }{ 38.99\% }
\newcommand{\claudeRIJavaPerfectLLPV }{ 43.30\% }
\newcommand{\claudegeminiIIflashJavaPerfectLLPV }{ 38.61\% }
\newcommand{\claudegeminiIproJavaPerfectLLPV }{ 40.66\% }
\newcommand{\claudeoIIIminiJavaPerfectLLPV }{ 40.76\% }
\newcommand{\RIqwenJavaPerfectLLPV }{ 31.69\% }
\newcommand{\geminiIIflashqwenJavaPerfectLLPV }{ 22.47\% }
\newcommand{\geminiIproqwenJavaPerfectLLPV }{ 29.64\% }
\newcommand{\oIIIminiqwenJavaPerfectLLPV }{ 33.19\% }
\newcommand{\RIgeminiIIflashJavaPerfectLLPV }{ 31.87\% }
\newcommand{\RIgeminiIproJavaPerfectLLPV }{ 37.81\% }
\newcommand{\oIIIminiRIJavaPerfectLLPV }{ 38.37\% }
\newcommand{\geminiIprogeminiIIflashJavaPerfectLLPV }{ 29.94\% }
\newcommand{\oIIIminigeminiIIflashJavaPerfectLLPV }{ 33.82\% }
\newcommand{\oIIIminigeminiIproJavaPerfectLLPV }{ 37.69\% }
\newcommand{\claudeqwenPHPPerfectLLPV }{ 23.35\% }
\newcommand{\RIclaudePHPPerfectLLPV }{ 28.36\% }
\newcommand{\claudegeminiIIflashPHPPerfectLLPV }{ 24.08\% }
\newcommand{\claudegeminiIproPHPPerfectLLPV }{ 26.77\% }
\newcommand{\oIIIminiclaudePHPPerfectLLPV }{ 24.88\% }
\newcommand{\RIqwenPHPPerfectLLPV }{ 27.48\% }
\newcommand{\geminiIIflashqwenPHPPerfectLLPV }{ 20.44\% }
\newcommand{\geminiIproqwenPHPPerfectLLPV }{ 23.30\% }
\newcommand{\oIIIminiqwenPHPPerfectLLPV }{ 24.22\% }
\newcommand{\RIgeminiIIflashPHPPerfectLLPV }{ 29.12\% }
\newcommand{\RIgeminiIproPHPPerfectLLPV }{ 30.93\% }
\newcommand{\RIoIIIminiPHPPerfectLLPV }{ 28.57\% }
\newcommand{\geminiIprogeminiIIflashPHPPerfectLLPV }{ 24.97\% }
\newcommand{\oIIIminigeminiIIflashPHPPerfectLLPV }{ 26.36\% }
\newcommand{\oIIIminigeminiIproPHPPerfectLLPV }{ 28.38\% }

\newcommand{\claudeqwenPythonTestPV }{ 40.53\% }
\newcommand{\RIclaudePythonTestPV }{ 65.48\% }
\newcommand{\geminiIIflashclaudePythonTestPV }{ 54.19\% }
\newcommand{\geminiIproclaudePythonTestPV }{ 56.56\% }
\newcommand{\oIIIminiclaudePythonTestPV }{ 43.66\% }
\newcommand{\RIqwenPythonTestPV }{ 61.34\% }
\newcommand{\geminiIIflashqwenPythonTestPV }{ 49.26\% }
\newcommand{\geminiIproqwenPythonTestPV }{ 48.39\% }
\newcommand{\oIIIminiqwenPythonTestPV }{ 46.29\% }
\newcommand{\RIgeminiIIflashPythonTestPV }{ 64.52\% }
\newcommand{\RIgeminiIproPythonTestPV }{ 63.85\% }
\newcommand{\RIoIIIminiPythonTestPV }{ 68.51\% }
\newcommand{\geminiIIflashgeminiIproPythonTestPV }{ 54.44\% }
\newcommand{\geminiIIflashoIIIminiPythonTestPV }{ 61.00\% }
\newcommand{\geminiIprooIIIminiPythonTestPV }{ 63.57\% }
\newcommand{\claudeqwenJavaTestPV }{ 55.14\% }
\newcommand{\claudeRIJavaTestPV }{ 59.39\% }
\newcommand{\claudegeminiIIflashJavaTestPV }{ 58.83\% }
\newcommand{\claudegeminiIproJavaTestPV }{ 56.93\% }
\newcommand{\oIIIminiclaudeJavaTestPV }{ 65.73\% }
\newcommand{\RIqwenJavaTestPV }{ 49.18\% }
\newcommand{\geminiIIflashqwenJavaTestPV }{ 41.02\% }
\newcommand{\geminiIproqwenJavaTestPV }{ 42.47\% }
\newcommand{\oIIIminiqwenJavaTestPV }{ 62.15\% }
\newcommand{\RIgeminiIIflashJavaTestPV }{ 54.25\% }
\newcommand{\RIgeminiIproJavaTestPV }{ 55.66\% }
\newcommand{\oIIIminiRIJavaTestPV }{ 62.31\% }
\newcommand{\geminiIprogeminiIIflashJavaTestPV }{ 46.38\% }
\newcommand{\oIIIminigeminiIIflashJavaTestPV }{ 63.20\% }
\newcommand{\oIIIminigeminiIproJavaTestPV }{ 63.58\% }
\newcommand{\claudeqwenPHPTestPV }{ 52.10\% }
\newcommand{\claudeRIPHPTestPV }{ 58.22\% }
\newcommand{\claudegeminiIIflashPHPTestPV }{ 51.46\% }
\newcommand{\claudegeminiIproPHPTestPV }{ 52.09\% }
\newcommand{\oIIIminiclaudePHPTestPV }{ 61.56\% }
\newcommand{\RIqwenPHPTestPV }{ 56.20\% }
\newcommand{\qwengeminiIIflashPHPTestPV }{ 43.35\% }
\newcommand{\geminiIproqwenPHPTestPV }{ 45.15\% }
\newcommand{\oIIIminiqwenPHPTestPV }{ 59.00\% }
\newcommand{\RIgeminiIIflashPHPTestPV }{ 55.35\% }
\newcommand{\RIgeminiIproPHPTestPV }{ 59.04\% }
\newcommand{\oIIIminiRIPHPTestPV }{ 63.12\% }
\newcommand{\geminiIprogeminiIIflashPHPTestPV }{ 44.55\% }
\newcommand{\oIIIminigeminiIIflashPHPTestPV }{ 59.58\% }
\newcommand{\oIIIminigeminiIproPHPTestPV }{ 59.82\% }

\newcommand{\claudeqwenPythonPerfectLLPV }{ 14.93\% }
\newcommand{\claudeRIPythonPerfectLLPV }{ 17.76\% }
\newcommand{\geminiIIflashclaudePythonPerfectLLPV }{ 24.57\% }
\newcommand{\claudegeminiIproPythonPerfectLLPV }{ 22.14\% }
\newcommand{\oIIIminiclaudePythonPerfectLLPV }{ 15.86\% }
\newcommand{\RIqwenPythonPerfectLLPV }{ 14.37\% }
\newcommand{\geminiIIflashqwenPythonPerfectLLPV }{ 21.59\% }
\newcommand{\geminiIproqwenPythonPerfectLLPV }{ 17.27\% }
\newcommand{\oIIIminiqwenPythonPerfectLLPV }{ 12.43\% }
\newcommand{\geminiIIflashRIPythonPerfectLLPV }{ 23.71\% }
\newcommand{\geminiIproRIPythonPerfectLLPV }{ 19.52\% }
\newcommand{\oIIIminiRIPythonPerfectLLPV }{ 15.66\% }
\newcommand{\geminiIIflashgeminiIproPythonPerfectLLPV }{ 24.75\% }
\newcommand{\geminiIIflashoIIIminiPythonPerfectLLPV }{ 25.81\% }
\newcommand{\geminiIprooIIIminiPythonPerfectLLPV }{ 21.72\% }

\newcommand{\claudeqwenPythonBASICPV }{ 12.97\% }
\newcommand{\claudeRIPythonBASICPV }{ 16.29\% }
\newcommand{\geminiIIflashclaudePythonBASICPV }{ 18.39\% }
\newcommand{\claudegeminiIproPythonBASICPV }{ 17.64\% }
\newcommand{\oIIIminiclaudePythonBASICPV }{ 17.87\% }
\newcommand{\RIqwenPythonBASICPV }{ 13.48\% }
\newcommand{\geminiIIflashqwenPythonBASICPV }{ 16.31\% }
\newcommand{\geminiIproqwenPythonBASICPV }{ 12.24\% }
\newcommand{\oIIIminiqwenPythonBASICPV }{ 14.46\% }
\newcommand{\geminiIIflashRIPythonBASICPV }{ 17.82\% }
\newcommand{\RIgeminiIproPythonBASICPV }{ 16.65\% }
\newcommand{\oIIIminiRIPythonBASICPV }{ 16.80\% }
\newcommand{\geminiIIflashgeminiIproPythonBASICPV }{ 17.35\% }
\newcommand{\oIIIminigeminiIIflashPythonBASICPV }{ 21.17\% }
\newcommand{\oIIIminigeminiIproPythonBASICPV }{ 17.73\% }

\newcommand{\claudeqwenJavaBASICPV }{ 29.79\% }
\newcommand{\claudeRIJavaBASICPV }{ 35.12\% }
\newcommand{\claudegeminiIIflashJavaBASICPV }{ 32.08\% }
\newcommand{\claudegeminiIproJavaBASICPV }{ 35.30\% }
\newcommand{\claudeoIIIminiJavaBASICPV }{ 31.98\% }
\newcommand{\RIqwenJavaBASICPV }{ 30.65\% }
\newcommand{\qwengeminiIIflashJavaBASICPV }{ 18.22\% }
\newcommand{\geminiIproqwenJavaBASICPV }{ 28.96\% }
\newcommand{\oIIIminiqwenJavaBASICPV }{ 24.19\% }
\newcommand{\RIgeminiIIflashJavaBASICPV }{ 29.95\% }
\newcommand{\geminiIproRIJavaBASICPV }{ 36.99\% }
\newcommand{\RIoIIIminiJavaBASICPV }{ 32.27\% }
\newcommand{\geminiIprogeminiIIflashJavaBASICPV }{ 27.86\% }
\newcommand{\oIIIminigeminiIIflashJavaBASICPV }{ 26.40\% }
\newcommand{\geminiIprooIIIminiJavaBASICPV }{ 30.87\% }

\newcommand{\claudeqwenPHPBASICPV }{ 24.92\% }
\newcommand{\RIclaudePHPBASICPV }{ 30.78\% }
\newcommand{\geminiIIflashclaudePHPBASICPV }{ 26.47\% }
\newcommand{\geminiIproclaudePHPBASICPV }{ 30.87\% }
\newcommand{\claudeoIIIminiPHPBASICPV }{ 23.94\% }
\newcommand{\RIqwenPHPBASICPV }{ 29.90\% }
\newcommand{\geminiIIflashqwenPHPBASICPV }{ 24.70\% }
\newcommand{\geminiIproqwenPHPBASICPV }{ 29.87\% }
\newcommand{\qwenoIIIminiPHPBASICPV }{ 21.55\% }
\newcommand{\RIgeminiIIflashPHPBASICPV }{ 32.14\% }
\newcommand{\geminiIproRIPHPBASICPV }{ 32.61\% }
\newcommand{\RIoIIIminiPHPBASICPV }{ 31.98\% }
\newcommand{\geminiIprogeminiIIflashPHPBASICPV }{ 30.24\% }
\newcommand{\geminiIIflashoIIIminiPHPBASICPV }{ 26.47\% }
\newcommand{\geminiIprooIIIminiPHPBASICPV }{ 30.31\% }

\newcommand{\averageMHTestPI }{ 27.50\% }
\newcommand{\averageMHTestPIII }{ 35.30\% }
\newcommand{\averageMHTestPV }{ 38.11\% }
\newcommand{\averageSHTestPI }{ 34.37\% }
\newcommand{\averageSHTestPIII }{ 42.60\% }
\newcommand{\averageSHTestPV }{ 45.98\% }
\newcommand{\averageSLTestPI }{ 45.06\% }
\newcommand{\averageSLTestPIII }{ 53.83\% }
\newcommand{\averageSLTestPV }{ 57.06\% }

\newcommand{\claudeCrossLanguageMeanBASICPI }{ 13.97\% }
\newcommand{\claudeCrossLanguageMeanBASICPIII }{ 20.83\% }
\newcommand{\claudeCrossLanguageMeanBASICPV }{ 24.06\% }
\newcommand{\claudeHaikuCrossLanguageMeanBASICPI }{ 13.24\% }
\newcommand{\claudeHaikuCrossLanguageMeanBASICPIII }{ 18.30\% }
\newcommand{\claudeHaikuCrossLanguageMeanBASICPV }{ 20.30\% }
\newcommand{\gemmaCrossLanguageMeanBASICPI }{ 6.02\% }
\newcommand{\gemmaCrossLanguageMeanBASICPIII }{ 12.59\% }
\newcommand{\gemmaCrossLanguageMeanBASICPV }{ 16.71\% }
\newcommand{\qwenCrossLanguageMeanBASICPI }{ 9.33\% }
\newcommand{\qwenCrossLanguageMeanBASICPIII }{ 13.11\% }
\newcommand{\qwenCrossLanguageMeanBASICPV }{ 14.88\% }
\newcommand{\cLlamaSCrossLanguageMeanBASICPI }{ 4.25\% }
\newcommand{\cLlamaSCrossLanguageMeanBASICPIII }{ 10.23\% }
\newcommand{\cLlamaSCrossLanguageMeanBASICPV }{ 14.40\% }
\newcommand{\LlamaIIICrossLanguageMeanBASICPI }{ 9.71\% }
\newcommand{\LlamaIIICrossLanguageMeanBASICPIII }{ 13.93\% }
\newcommand{\LlamaIIICrossLanguageMeanBASICPV }{ 15.85\% }
\newcommand{\CLlamaLCrossLanguageMeanBASICPI }{ 1.69\% }
\newcommand{\CLlamaLCrossLanguageMeanBASICPIII }{ 4.26\% }
\newcommand{\CLlamaLCrossLanguageMeanBASICPV }{ 6.27\% }
\newcommand{\DSCoderCrossLanguageMeanBASICPI }{ 5.32\% }
\newcommand{\DSCoderCrossLanguageMeanBASICPIII }{ 10.41\% }
\newcommand{\DSCoderCrossLanguageMeanBASICPV }{ 13.47\% }
\newcommand{\RICrossLanguageMeanBASICPI }{ 13.24\% }
\newcommand{\RICrossLanguageMeanBASICPIII }{ 21.55\% }
\newcommand{\RICrossLanguageMeanBASICPV }{ 25.85\% }
\newcommand{\geminiIIflashCrossLanguageMeanBASICPI }{ 14.21\% }
\newcommand{\geminiIIflashCrossLanguageMeanBASICPIII }{ 19.04\% }
\newcommand{\geminiIIflashCrossLanguageMeanBASICPV }{ 21.11\% }
\newcommand{\geminiIproCrossLanguageMeanBASICPI }{ 14.04\% }
\newcommand{\geminiIproCrossLanguageMeanBASICPIII }{ 20.85\% }
\newcommand{\geminiIproCrossLanguageMeanBASICPV }{ 23.71\% }
\newcommand{\gptIVoCrossLanguageMeanBASICPI }{ 11.41\% }
\newcommand{\gptIVoCrossLanguageMeanBASICPIII }{ 18.65\% }
\newcommand{\gptIVoCrossLanguageMeanBASICPV }{ 22.11\% }
\newcommand{\oIIIminiCrossLanguageMeanBASICPI }{ 13.13\% }
\newcommand{\oIIIminiCrossLanguageMeanBASICPIII }{ 17.36\% }
\newcommand{\oIIIminiCrossLanguageMeanBASICPV }{ 19.36\% }

%% file: discussion.tex
\section{Discussion \& Threats to Validity}
%


\paragraph*{Comparison to Agentic Workflows}
Recent research in APR~\cite{autocoderover,specrover,yang2024swe} focused on building agentic workflows, which is not part of our empirical exploration. Instead, we focused on a more fundamental evaluation of LLMs for the under-explored area of cross-language repair capabilities. Compared to agentic workflows, we focus on single prompt setups using different prompt ingredients. Nevertheless, our insights will help other researchers and engineers select LLMs to design and build new agentic workflows that can handle cross-language repair scenarios.

\paragraph*{Experiments with LLMs}
Sallou et al.~\cite{DBLP:conf/icse/SallouDP24} discussed several evaluation aspects when using LLMs in software engineering research, including the problem of output variability, data leakage, and closed-source models.
To address the output variability of LLMs, we generated not only one but 15 patches (see details in Section~\ref{sec:metrics}). To evaluate the generated patches, we used the established $pass@k$ metric~\cite{DBLP:journals/corr/abs-2107-03374} representing an estimator for the model's effectiveness. Assessing only the plausibility (i.e., passing test cases) and not the correctness (i.e., semantical equivalence with a ground truth like a developer patch) represents a threat to construct validity. However, manually assessing the actual correctness for all subjects is not feasible.
The data leakage issue is more challenging to address because we cannot control the training data of state-of-the-art LLMs. Related works, e.g., RepairBench~\cite{DBLP:journals/corr/abs-2409-18952} used a dataset with more recent data, whose time period is now also included in the cut-off dates of the latest models. At this point, we have to assume that some of the benchmark data may have been included in the training corpora of some of the LLMs we used in our experiments, which represents a threat to internal validity.
In fact, Zhou et al.~\cite{zhou2025lessleakbenchinvestigationdataleakage} recently investigated 83 SE benchmarks regarding their potential data leakage. According to their analysis \defectsFourJ (in its 2.0 version) has a leaked ratio of 0.41\% and BugsInPy 11.0\%. Both numbers are relatively low. Interestingly, although BugsInPy obviously has some data leakage issues, our experiments showed it as the most challenging benchmark for the tested LLMs.
Ramos et al.~\cite{RamosMJCGG25} further studied the impact of benchmark leakage across several open-source models and found that, although some models show substancial evidence of memorization on widely used benchmarks such as \defectsFourJ, newer models, trained on larger datasets such as Llama 3.1 show only limited signs of leakage.
%
%
%
%
%
As part of our empirical evaluation, we support a comparative analysis of models by selecting a wide range of open-source/weight and closed models. Our broad selection of models and benchmarks reduces the threat to external validity; however, we cannot claim generality beyond our experiments.

\paragraph*{Prompt Optimization}
As described in Section~\ref{sec:prompt-design}, we used zero-shot prompts with different repair ingredients in our experiments. Those prompts were designed based on and in consistency with prior work~\cite{DBLP:journals/corr/abs-2404-05520,silva2024repairllama,Xia2023,DBLP:journals/corr/abs-2409-18952}, and have been tested in preliminary experiments. Additional optimization and tuning of the prompts could lead to better repair performances on our benchmarks; however, we did intentionally not perform further optimizations to prevent introducing model-specific biases and further benchmark overfitting.

\paragraph*{Repair Time}
We noticed that, e.g., due to its reasoning strategy, the \deepseekROneDist model produces more iterative output, requiring significantly more time to conclude with the final repair than the other models. Other studies~\cite{Noller2022} showed that developers could accept a timeout of up to one hour for integrating APR technology in their workflows. This timeout is still much higher than LLMs would require; in our experiments spanning between seconds and several minutes. However, it is currently unknown what an acceptable timeout for LLM-based APR would be so that, e.g., it can effectively integrate into agentic workflows. Different resource demands across the models, in particular with respect to timing, can lead to unfair comparison representing a threat to internal validity. Our study methodology did not limit the time for model inference or patch validation because we wanted to focus on the actual repair capabilities, and we leave a more rigorous analysis regarding repair time for future work.


%% file: conclusion.tex
\section{Conclusion} 
This work demonstrates an intensive exploration of the current capabilities of state-of-the-art LLMs to perform generalizable repair, i.e., across multiple languages and patch complexities. After comparing open models with closed models on benchmarks for Java, JavaScript, PHP, and Python, we conclude that no single model outperforms the others on all benchmarks. Instead, it becomes necessary to combine various models to realize cross-language repair capabilities. In our experiments with various prompting strategies, we observed that LLM-based APR with automated/imperfect fault localization leads to significantly lower repair performance. Therefore, we advocate for more realistic evaluations in the APR community.
In the future, we plan to extend our study to more aspects, such as repair time, particularly regarding reasoning models. Further, we want to understand closer the challenges in repairing across language barriers and how to detect uncertainties of models in code/patch generation automatically.
We believe that the insights of this study are crucial to further innovate the integration of APR technologies in software development practice.

%% file: disclaimer.tex
\section{Disclaimer}
The views and conclusions expressed in this paper are those of the authors alone and do not represent the official policies or endorsements of SonarSource or any of its subsidiaries or affiliates. Furthermore, the findings presented herein are independent and should not be interpreted as an evaluation of the quality of SonarSource’s products.